\documentclass[a4paper,12pt]{article}
\pdfoutput=1
\usepackage{amsmath}
\usepackage{amssymb}
\usepackage{amsfonts,times}
\usepackage{theorem}
\usepackage{hyperref}
\usepackage{graphicx}
\usepackage{caption}
\usepackage{longtable}
\usepackage{multirow}
\usepackage{multicol}
\usepackage{placeins}
\usepackage{array}
\AtBeginDocument{}
\newtheorem{theorem}{Theorem}

\newtheorem{definition}[theorem]{Definition}

{\theorembodyfont{\upshape}}
{\theorembodyfont{\upshape}}

\newcommand{\cC}{\mathcal{C}}

\newcommand{\cL}{\mathcal{L}}

\newcommand{\cN}{\mathcal{N}}

\newcommand{\cP}{\mathcal{P}}
\newcommand{\cW}{\mathcal{W}}
\newcommand{\cV}{\mathcal{V}}
\newcommand{\cD}{\mathcal{D}}
\newcommand{\cS}{\mathcal{S}}
\newcommand{\cU}{\mathcal{U}}

\newcommand{\sumBinoms}{\genfrac{\{}{\}}{0pt}{}}
\begin{document}

\title{Double Majority and Generalized Brexit: Explaining Counterintuitive Results}
\author{Werner Kirsch \footnote{Fakult\"{a}t f\"{u}r Mathematik und Informatik, FernUniversit\"{a}t Hagen, Germany.}, Wojciech S{\l}omczy\'{n}ski
\footnote{Jagiellonian Center for Quantitative Research in Political Science / Institute of Mathematics, Jagiellonian University, Cracow, Poland.}, \\ Dariusz Stolicki \footnote{Jagiellonian Center for Quantitative Research in Political Science / Institute of Political Science and International Relations, Jagiellonian University, Cracow, Poland.}, Karol \.{Z}yczkowski\footnote{Jagiellonian Center for Quantitative Research in Political Science / Institute of Physics, Jagiellonian University, Cracow, Poland.}}

\date{December 14, 2018}

\maketitle
\begin{abstract}
A mathematical analysis of the distribution of voting power in the Council of the European Union operating according to the Treaty of Lisbon is presented. We study the effects of Brexit on the voting power of the remaining members, measured by the Penrose--Banzhaf Index. We note that the effects in question are non-monotonic with respect to voting weights, in fact, some member states will \emph{lose} power after Brexit.
We use the normal approximation of the Penrose--Banzhaf Index in double-majority games to show that such non-monotonicity is in most cases inherent in the double-majority system, but is strongly exacerbated by the peculiarities of the EU population vector. Furthermore, we investigate consequences of a hypothetical "generalized Brexit", i.e., {\sl NN-exit} of another member state (from a 28-member Union), noting that the effects on voting power are non-monotonic in most cases, but strongly depend on the size of the country leaving the Union.
 \end{abstract}

\section{Introduction}
The voting rules for the Council of the European Union are based on the Treaty
of Lisbon. A decision of the Council about a proposal of the Commission requires
a `double majority': A proposal is approved if $55\%$ of the member states support
it which also represent $65\%$ of the Union's population.  Formally speaking,
this is a union of two weighted voting systems. In the first subsystem every
country has weight $1$ and the relative quota is $55\%$ (thus the absolute quota
is $16$ before and $15$ after Brexit). In the second subsystem the weights are
given by the population of the respective country and the relative quota is
$65\%$ (for more details see the next section below). There is also a third voting system 
involved: A proposal is also if approved if less than four members object, even if the population criterion
is violated. However, this `third rule' of the `double majority' plays only a marginal role, as we explain
in more detail below (see Subsection \ref{subsV}).

Intuitively, it seems to be clear that after Brexit the influence of each state
in the Council (except UK, of course) should grow as the normalized weight
increases for both subsystems. It was observed independently in \cite{Kirsch1},
\cite{Koczy}, \cite{Gollner}, \cite{Szcz}, and \cite{Petroczy} that this is \emph{not} the case.
The power as defined by the Banzhaf index grows indeed for all bigger and medium
size states. However, the seven smallest states \emph{lose} power through Brexit.
While this fact has been noted in earlier works, we move beyond observation and
seek to explain it. First, we analyze how this effect may be triggered by the
double majority principle, by decomposing the two sources of voting power
arising from the two subsystems described below. Second, we consider
"generalized Brexits" (N.N.-Exits), i.e., exits of other current member states,
and analyzing the ratio of post-exit to pre-exit voting power for any remaining
country as a function of population. On the basis of such analysis, we
distinguish between three patterns of N.N.-Exit effects and discuss how this
effect may result from the relationship between the distribution of population
within the EU and the qualified majority quota.

\section{Framework and Tools}

\begin{definition}
   A \emph{voting system} consist of a (finite) set $V$ of voters and a set $\cW\subset \cP(V)$ of winning coalitions, satisfying
   \begin{enumerate}
      \item $V\in\cW$
      \item $\emptyset\not\in\cW$
      \item If $A\in\cW$ and $A\subset B\subset V$ then $B\in\cW$
         \end{enumerate}

  In a \emph{weighted voting system} with \emph{weights} $w_v\geq 0$ for each $v\in V$ and \emph{quota} $q$ the set of winning coalitions is given by
  \begin{align}
     \cW~=~\{A\subset V\mid \sum_{v\in A}w_v\geq q\}
  \end{align}
  We set $w(A)=\sum_{v\in A}w_{v}$ and call the number $r=\frac{q}{w_v}$ the \emph{relative quota}.

  We denote a weighted voting system with weight $w_{i},i=1,\ldots,N$ and (absolute) quota $q$ by
  $[q;w_{1},\ldots,w_{N}]$.
\end{definition}
\begin{definition}
   A voter $v$ is called \emph{decisive} for a coalition $A\subset V$ if either $v\in A, A\in\cW$ and $A\setminus\{v\}\not\in\cW$
   or $v\not\in A, A\not\in\cW$ and $A\cup\{v\}\in\cW$. The set of coalitions for which $v$ is decisive is denoted by $\cD(v)$.

   The \emph{Banzhaf Power} $\psi_v$ of a voter $v$ is defined by
   \begin{align}
      \psi_v~:=~\frac{\#\{ A\mid A\in\cD(v) \}}{2^{\#V}}\,,
   \end{align}
   where $\#A $ is the number of elements in A.

   The \emph{Banzhaf Index} $\beta_v$ \cite{Penrose, Banzhaf} is the `relative' Banzhaf Power defined as
   \begin{align}
      \beta_v~:=~\frac{\psi(v)}{\sum_{w\in V}\psi_w}.
   \end{align}
\end{definition}
\begin{definition}
   The \emph{Shapley-Shubik Index} counts the number of \emph{permutations} for which $v$ is decisive.
   A permutation of a (finite) set $V$ is an ordering of the elements of $V$.
   If $V$ has $N$ elements and $\pi=v_{1}, v_{2},\ldots,v_{N} $ is a permutation of $V$
   then the voter $v_{k}$ is called decisive (or pivotal) for $\pi$ if $\{ v_{1},\ldots,v_{k} \}\in\cW $,
   but $\{ v_{1},\ldots,v_{k-1} \}\not\in\cW $. We denote the set of all permutations of $V$ by $\cS(V)$ and
   the set of permutations for which $v$ is decisive by $\cS_{v}(V)$.

   The \emph{Shapley-Shubik Index} $S(v)$ of a voter $v$ is defined by
   \begin{align}
      S(v)~=~\frac{\#\cS_{v}(V)}{\#\cS(V)}
   \end{align}
\end{definition}

Both the Shapley-Shubik Index and the Banzhaf Index measure the power of voters in a voting system. Their difference lies in the assumed collective behavior of the voters (see e.\,g. \cite{KSurv}).

\section{Theoretical models of exit effects}

\subsection{General considerations}

In this paper we investigate how the power structure is changed if a voter leaves the voting system.
Given a voting system $(V,\cW)$ and a voter $v_{0}\in V$ who leaves the system we have to determine the voting rules for
the set $V'=V\setminus\{ v_{0} \}$ of remaining voters.

For a weighted voting system it is natural to keep the weights for the remaining voters. It is perhaps less obvious
what to do with the quota.

Suppose we start with a weighted voting system $\cV=[q,w_{1},\ldots,w_{N}]$ from which voter $N$ defects then
the voting system is $\cV'=[q',w_{1},\ldots,w_{N-1}]$.

There seem to be three reasonable ways to determine the new quota: The first is to fix
the \emph{relative} quota, another way is to fix the \emph{absolute} quota, yet another to fix the difference between the
total weights and the quota.

This motivates the following definition.
\begin{definition}\label{def:quotas}
   Suppose $\cV=[q,w_{1},\ldots,w_{N}]$ is a weighted voting system and set $W=w_v$ and $W'=w_v-w_{N}$.

   We define the following
   weighted voting systems for the set $V'=\{ v_{1},\ldots,v_{N-1} \}$ of voters
   \begin{enumerate}
      \item The weighted voting system $\tilde{\cV}$ with fixed relative quota
             \begin{align}
                \tilde{\cV}~=~[\tilde{q},w_{1},\ldots,w_{N-1}]\qquad \text{with }\quad\tilde{q}=\frac{W'}{W}\,q
             \end{align}
             \item The weighted voting system $\overline{\cV}$ with fixed absolute quota
             \begin{align}
                \overline{\cV}~=~[\overline{q},w_{1},\ldots,w_{N-1}]\qquad \text{with }\quad\overline{q}=q
             \end{align}
             provided $q<W'$.
             \item The weighted voting system $\underline{\cV}$ with fixed difference to the total weight
             \begin{align}
                \underline{\cV}~=~[\underline{q},w_{1},\ldots,w_{N-1}]\qquad \text{with }\quad\underline{q}=q-(W-W')
             \end{align}
             provided $q>W-W'$.
   \end{enumerate}
\end{definition}

Intuitively, one is tempted to expect that if one voter leaves the voting system, the power of each other voter should \emph{increase}. However, this is not the case, in general.

For example in the weighted voting system $\cV=[3;3,1,1,1] $ each voter has positive power, for example the voters with weight $1$
have $\beta(v)=\frac{1}{10}$ and $S(v)=\frac{1}{12}$. If the last player defects, the other small players become completely powerless
regardless of which of the quotas in Definition \ref{def:quotas} is used.

If a weighted voting system with $N$ voters is simple. i.\,e. if all weights are equal, then both the Banzhaf- and the Shapley-Shubik-Index
equal $\frac{1}{N}$, so they are \emph{increasing} if voters leave the system.

\subsection{Jagiellonian compromise}

On the basis of Penrose's work \cite{Penrose} several authors (e.\,g. \cite{KirschZeit}, \cite{KirschHomoE}, \cite{SlomZycz}) suggested that the weights or rather the power indices of the countries in the Council should be proportional to the \emph{square root} of the population of the respective country. Such an idea was applied in the voting system known as the Jagiellonian Compromise \cite{SlomZycz}, which gives every member state a voting weight proportional to the square root of its population $P_{i}$ \emph{and} sets the quota to
\begin{align}
q~=~\frac{1}{2}\,\Big(1+\frac{\sqrt{\sum_{i=1}^N  P_{i}}}{\sum_{i=1}^N 
\sqrt{P_{i}}}\,\Big)\,.
\end{align}
This threshold minimizes the distance between the Banzhaf indices of all member states and their respective voting weights.

\section{Lisbon treaty and the Brexit}

\subsection{Voting in the Council}\label{subsV}

The treaty of Lisbon stipulates a complex voting system for the Council of the EU. A proposal of the Commission is approved by the Council if:\\
\emph{Either} at least $q_1 := 55\%$ of the member states support the proposal and they represent at least $q_2 := 65\%$ of the Union's population \emph{or}
all but at most $3$ vote `yea'.

If we denote by $P_{1},P_{2},\ldots,P_{N}$ the population of $N$ member states and $P=\sum P_{i}$ the population of the Union, then the voting
system in the Council is a combination of the following weighted voting systems
\begin{align}\label{eu1}
   \cV_{N}^{1}~&=~[65\cdot P;P_{1},\ldots,P_{N}]\\
   \cV_{N}^{2}~&=~[55\cdot N;1,1,\ldots,1]\label{eu2}\\
   \text{and}\quad\cV_{N}^{3}~&=~[N-3;1,1,\ldots,1]\label{eu3}
\end{align}

The voting system for the Council is given by
\begin{align}
   \cV_{N}~=~\Big(\cV_{N}^{1}\cap\cV_{N}^{2}\Big)\ \cup\ \cV_{N}^{3}
\end{align}

Recall that a coalition $C$ in $\cU\cap\cW$ (resp. in $\cU\cup\cW$) is winning if $C$ is winning in  $\cU$ \emph{and} in $\cW$ (resp. winning in  $\cU$ \emph{or} in $\cW$).

The voting system $\cV_{28}$ (the system with the $28$ member states as of 2018) is \emph{not} a weighted system. As any voting system it can be obtained as an intersection of $D$ weighted voting systems for some $D$ (see e.\,g. \cite{TaylorP}). The smallest such $D$ is called the \emph{dimension} of the voting system. The dimension of $\cV_{EU}$ is at least $7$ \cite{KurzN}.

The voting systems $\cV_{N}^{1}$ and $\cV_{N}^{2}$ are systems with fixed relative quota as defined in Definition \ref{def:quotas}, while $\cV_{N}^{3}$ is a system with fixed difference to the total weight. The total system is therefore a `hybrid' system with respect to defection.

It is certainly a rather complicated voting system and we will see that it shows some rather unexpected results.

For practical purposes, however, this system can be somewhat simplified, as the effect of the third voting subsystem can be considered negligible. Under the current distribution of weights in the EU, there are $26 472 389$ quasi-minimal  winning coalitions (i.e., coalitions with at least one pivotal voter) in the $\cV_{28}^{1}\cap\cV_{28}^{2}$ voting system. Out of all coalitions winning under $\cV_{28}^{3}$, none can be losing under $\cV_{28}^{2}$, and only $10$ are losing under $\cV_{28}^{1}$, so the union of $\cV_{28}^{1}\cap\cV_{28}^{2}$ with $\cV_{28}^{3}$ only changes the status of those $10$ coalitions. 
The effect omitting $\cV_{28}^{3}$ is `biggest' for the smallest state, Malta. Even for Malta the omission of $\cV_{28}^{3}$ changes the voting power by about $2.51\cdot 10^{-6}$.
We will denote the pure double-majority system $\cV_{28}^{1}\cap\cV_{28}^{2}$ as $\cV_{28}^{*}$.

In the next section we'll analyze the effect of Brexit on the distribution of voting power among the remaining member states both under the Lisbon system and some of its variations and under the Jagiellonian Compromise.

\subsection{Brexit}

One would expect that after Brexit the voting power of the remaining countries should increase as the share of votes increases in all three subsystems $\cV^{1}, \cV^{2}, \cV^{3}$.

The increase in power is obvious for subsystems $\cV^{2}$ and $\cV^{3}$ since for these systems the voting weights are equal for all countries. A computation for the system $\cV^{1}$ shows monotonicity as well (see Table \ref{tab1}). This table shows the Banzhaf Indices for the EU Countries under the weighted voting system $\cV^{1}$ before and after Brexit. The column `relative difference' shows the quantity $\frac{\beta_i^{27}}{\beta_i^{28}}-1$, where $\beta_i^{27}$ is the pre-Brexit voting power and $\beta_i^{28}$ is the post-Brexit voting power. We will denote $\frac{\beta_i^{27}}{\beta_i^{28}}-1$ as $\varphi_i$.

\begin{figure}[h]
	\centering
	\includegraphics[width=\linewidth]{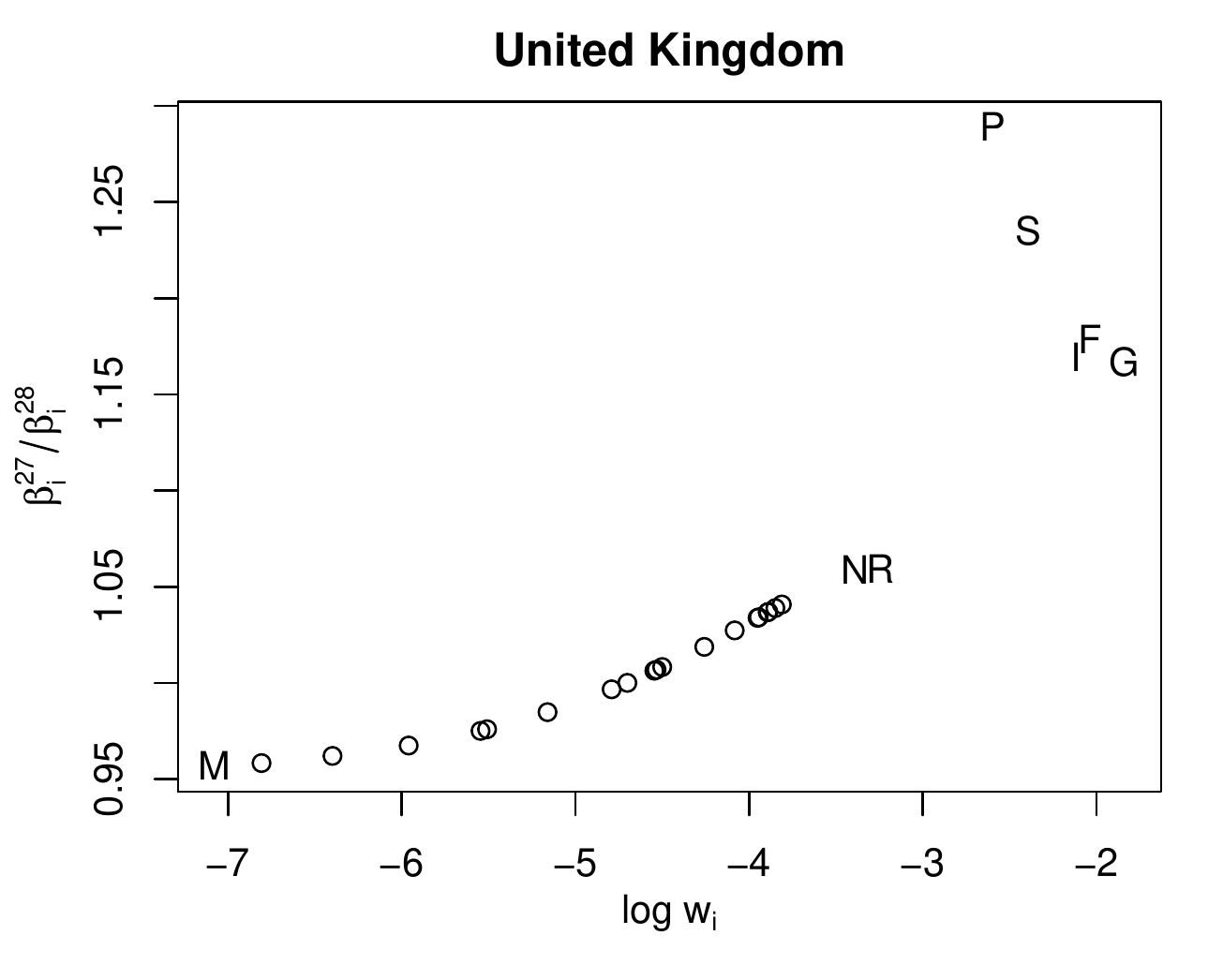}
	\caption{Effects of Brexit on the voting power: ratio of post-Brexit to pre-Brexit voting power (measured by normalized Banzhaf indices $\beta_i$) as a function of pre-Brexit voting weight $w_i$. Weights are on a logarithmic scale. Largest and smallest countries have been identified by the first letters of their names.}
	\label{fig:brexit}
\end{figure}

In contrast to its subsystems, the total system $\cV$ shows the remarkable effect that the eight smallest countries actually lose power (as measured by the Banzhaf Index) after Brexit (see Table \ref{tab2} and Fig. \ref{fig:brexit}). We note that the largest countries gain voting power as expected, but the gain (apart from a small anomaly arising for Italy, which may be a result of a numerical artifact) decreases monotonically with the country's size. From the perspective of voting power, Poland ($\log w_i \approx -2.59$) appears as the chief beneficiary of Brexit (gaining more than $28\%$ in terms of relative increase of power). Between Poland and the next-largest country, Romania ($\log w_i \approx -3.24$) an apparent discontinuity appears, and from Romania downward in population size the gains become smaller monotonically, ending with a loss of more than $4\%$ for Malta. The comparison of the Shapley-Shubik Indices show a somewhat different picture. The four smallest states again lose power, but the change in voting power is strictly increasing with the voting weight.

As we remarked already the system $\cV^{1}$ keeps the \emph{relative} quota fixed. In the case of Brexit this means that the absolute quota jumps from $16$ to $15$. If instead in the system $\cV^{1} $ we keep the \emph{absolute} quota fixed (at $q=16$) then the large countries lose power (see Table \ref{tab3}) and the smaller countries gain power. The same thing would happen if another country would leave the Union after Brexit (since absolute quota would remain fixed, as $\left \lceil{0.55*26}\right \rceil = \left \lceil{0.55*27}\right \rceil$).

It is interesting to compare these results with the `Jagiellonian Compromise' as voting system. For this system all countries win power
after Brexit as one should expect (see Table \ref{tab4}. In this case all countries gain power and the gain is uniform up to small variations.

The case that Scotland might separate from the United Kingdom after Brexit and join the European Union was considered in the paper \cite{Kirsch1}. Through Scotland's (hypothetical) joining the Union the bigger states will lose power while the smaller states win influence (see Table \ref{tab5}).

\subsection{Explaining nonmonotonicity: decomposing voter power in double-majority systems}

Under a double-majority system, such as $\cV_{EU}^{*}$, we can introduce additional measures of voting power that enable us to better understand the effects of both voting rules on each voter's power.

\begin{definition}
A set of coalitions which include voter $v$ ($\cC(v)$) can be partitioned into three sets:
\begin{itemize}
	\item $\cL(v)$ -- losing coalitions,
	\item $\cW_0(v)$ -- winning coalitions for which $v$ is non-pivotal,
	\item $\cW_1(v)$ -- winning coalitions for which $v$ is pivotal under $\cV_{N}^{1}$, but not under $\cV_{EU}^{2}$,
	\item $\cW_2(v)$ -- winning coalitions for which $v$ is pivotal under $\cV_{N}^{2}$, but not under $\cV_{EU}^{1}$,
	\item $\cW_3(v)$ -- winning coalitions for which $v$ is pivotal under both $\cV_{N}^{1}$ and $\cV_{EU}^{2}$.
\end{itemize}
We will denote the sum of $\cW_1(v)$, $\cW_2(v)$, and $\cW_3(v)$ by $\cW(v)$.
\end{definition}

Let us now forgo normalization for a while and just consider how the cardinalities of $\cL(v)$, $\cW_0(v)$, $\cW_1(v)$, $\cW_2(v)$, and $\cW_3(v)$ change when a member country $x$ leaves the Union. Let us denote the pre-exit coalitions by the superscript index $28$, and post-exit coalitions by the superscript index $27$. Finally, let
$$\Delta(v) := 2 \big(\#\cW_1^{27}(v) + \#\cW_2^{27}(v) + \#\cW_3^{27}(v)\big) - \big(\#\cW_1^{28}(v) + \#\cW_2^{28}(v) + \#\cW_3^{28}(v)\big).$$

Note that $1 + \Delta(v) / (b^{28}(v) 2^{\#V})$ (where the denominator is the pre-exit number of pivotal coalitions) is the ratio of the post-exit to pre-exit Banzhaf power. As it differs from the ratio of Banzhaf indices only as to a constant factor ($2(\#\cD^{27})/(\#\cD^{28})$, which equals approximately 1.051632 for Brexit), explaining the differences in $\Delta(v)$ among member states appears to be the key step in explaining the Brexit effects.

Let $c \in \cC^{28}(v)$, and $c' := c \setminus x$. Let us consider under what conditions post-exit coalition $c'$ can be of a different class than pre-exit coalition $c$. There are fifty combinations, some of which can be easily shown to be impossible. We analyze them in detail in Appendix B and summarize those which are possible in Table \ref{tab:ranges}, noting how each change in coalition status affects pivotality:

\bgroup
\def\arraystretch{1.5}
	\centering
	\addtolength{\LTleft} {-1.3in}
	\addtolength{\LTright}{-1.3in}
\begin{longtable}{|c|c|c|c|c|c|}
	\hline
	\textbf{$x \in c\,$} & \textbf{$\#c$} & \multicolumn{2}{c|}{\textbf{weight range}} & \textbf{pivot} & \textbf{change} \\
	\hline
	$x \notin c$ & $0-14$ & $0$ & $1$ & $0$ & $\cL^{28}$ to $\cL^{27}$ \\
	\hline
	$x \in c$ & $0-15$ & $0$ & $1$ & $0$ & $\cL^{28}$ to $\cL^{27}$ \\
	\hline \hline
	$x \notin c$ & $15$ & $0$ & $q_2 (1 - w_x)$ & $0$ & $\cL^{28}$ to $\cL^{27}$ \\
	\hline
	$x \notin c$ & $15$ & $q_2 (1 - w_x)$ & $\min\{{q_2 (1 - w_x) + w_v \atop q_2}\}$ & $1$ & $\cL^{28}$ to $\cW_3^{27}$ \\
	\hline
	$x \notin c$ & $15$ & $q_2 (1 - w_x) + w_v$ & $1$ & $1$ & $\cL^{28}$ to $\cW_1^{27}$ \\
	\hline \hline
	$x \notin c$ & $16-27$ & $0$ & $q_2 (1 - w_x)$ & $0$ & $\cL^{28}$ to $\cL^{27}$ \\
	\hline
	$x \notin c$ & $16-27$ & $q_2 (1 - w_x)$ & $\min\{{q_2 (1 - w_x) + w_v \atop q_2}\}$ & $1$ & $\cL^{28}$ to $\cW_2^{27}$ \\
	\hline
	$x \notin c$ & $16-27$ & $q_2 (1 - w_x) + w_v$ & $q_2$ & $0$ & $\cL^{28}$ to $\cW_0^{27}$ \\
	\hline
	$x \notin c$ & $16-27$ & $q_2$ & $q_2 (1 - w_x) + w_v$ & $0$ & $\cW_{2,3}^{28}$ to $\cW_2^{27}$ \\
	\hline
	$x \notin c$ & $16-27$ & $\max\{{q_2 (1 - w_x) + w_v \atop q_2}\}$ & $q_2 + w_v$ & $-1$ & $\cW_{2,3}^{28}$ to $\cW_0^{27}$ \\
	\hline
	$x \notin c$ & $16$ & $q_2 + w_v$ & $1$ & $-1$ & $\cW_1^{28}$ to $\cW_0^{27}$ \\
	\hline
	$x \notin c$ & $17-27$ & $q_2 + w_v$ & $1$ & $0$ & $\cW_0^{28}$ to $\cW_0^{27}$ \\
	\hline \hline
	$x \in c$ & $16$ & $0$ & $q_2$ & $0$ & $\cL^{28}$ to $\cL^{27}$ \\
	\hline
	$x \in c$ & $16$ & $q_2$ & $\min\{{q_2 (1 - w_x) + w_x \atop q_2 + w_v}\}$ & $-1$ & $\cW_3^{28}$ to $\cL^{27}$ \\
	\hline
	$x \in c$ & $16$ & $q_2 (1 - w_x) + w_x$ & $q_2 + w_v$ & $0$ & $\cW_3^{28}$ to $\cW_3^{27}$ \\
	\hline
	$x \in c$ & $16$ & $q_2 + w_v$ & $q_2 + (1 - q_2) w_x$ & $-1$ & $\cW_1^{28}$ to $\cL^{27}$ \\
	\hline
	$x \in c$ & $16$ & $\max\{{q_2 + (1 - q_2) w_x \atop q_2 + w_v}\}$ & $q_2 + (1 - q_2) w_x + w_v$ & $0$ & $\cW_1^{28}$ to $\cW_3^{27}$ \\
	\hline
	$x \in c$ & $16$ & $q_2 + (1 - q_2) w_x + w_v$ & $1$ & $0$ & $\cW_1^{28}$ to $\cW_1^{27}$ \\
	\hline \hline
	$x \in c$ & $17-28$ & $0$ & $q_2$ & $0$ & $\cL^{28}$ to $\cL^{27}$ \\
	\hline
	$x \in c$ & $17-28$ & $q_2$ & $\min\{{q_2 + (1 - q_2) w_x \atop q_2 + w_v}\}$ & $-1$ & $\cW_2^{28}$ to $\cL^{27}$ \\
	\hline
	$x \in c$ & $17-28$ & $q_2 + (1 - q_2) w_x$ & $q_2 + w_v$ & $0$ & $\cW_2^{28}$ to $\cW_2^{27}$ \\
	\hline
	$x \in c$ & $17-28$ & $q_2 + w_v$ & $q_2 (1 - w_x) + w_x$ & $0$ & $\cW_0^{28}$ to $\cL^{27}$ \\
	\hline
	$x \in c$ & $17-28$ & $\max\{{q_2 + (1 - q_2) w_x \atop q_2 + w_v}\}$ & $q_2 + (1 - q_2) w_x + w_v$ & $1$ & $\cW_0^{28}$ to $\cW_2^{27}$ \\
	\hline
	$x \in c$ & $17-28$ & $q_2 + (1 - q_2) w_x + w_v$ & $1$ & $0$ & $\cW_0^{28}$ to $\cW_0^{27}$ \\
	\hline
\caption{Coalition status changes for country $v$ after country $x$ exits the EU of 28 states and their effect on each coalition's pivotality ($1$ means a change from non-pivotal to pivotal, $-1$ -- from pivotal to non-pivotal, and $0$ -- no change in pivotality)}\label{tab:ranges}
\end{longtable}
\egroup

It follows that the change of the Banzhaf Power of voter $v$ after the exit of voter $x$ can be expressed as:
\allowdisplaybreaks
\begin{small}
\begin{align} \label{banChange}
\frac{\Delta(v)}{2^n} & = \psi^{N}(v) - 2 \psi^{N-1}(v) = \\
& \Pr\bigg(w(c) \in \Big(q_2 (1 - w_x), \min\Big\{{q_2 (1 - w_x) + w_v \atop q_2}\Big\}\Big) , {x \notin c \atop \#c \ge K-1}\bigg) \label{banChange1} \\
& +\Pr\bigg(w(c) \in \Big(q_2 (1 - w_x) + w_v, 1\Big) , {x \notin c \atop \#c = K-1}\bigg) \\
& -\Pr\bigg(w(c) \in \Big(\max\Big\{{q_2 (1 - w_x) + w_v \atop q_2}\Big\}, q_2 + w_v\Big) , {x \notin c \atop \#c \ge K}\bigg) \label{banChange3} \\
& -\Pr\bigg(w(c) \in \Big(q_2 + w_v, 1\Big) , {x \notin c \atop \#c = K}\bigg) \\
& -\Pr\bigg(w(c) \in \Big(q_2, \min\Big\{{q_2 + (1 - q_2) w_x \atop q_2 + w_v}\Big\}\Big) , {x \in c \atop \#c \ge K}\bigg) \label{banChange5} \\
& -\Pr\bigg(w(c) \in \Big(q_2 + w_v, q_2 + (1 - q_2) w_x\Big) , {x \in c \atop \#c = K}\bigg) \\
& +\Pr\bigg(w(c) \in \Big(\max\Big\{{q_2 + (1 - q_2) w_x \atop q_2 + w_v}\Big\}, q_2 + (1 - q_2) w_x + w_v\Big) , {x \in c \atop \#c > K}\bigg), \label{banChange7}
\end{align}
\end{small}
where $N = 28$ and $K := \lceil q_1 N \rceil = 16$. Note again that this formula is only correct under the assumption that $\lceil q_1 N \rceil > \lceil q_1 (N-1) \rceil$.

Starting with Merrill \cite{Merrill}, researchers have approximated the distribution of weights for all coalitions (regardless of size) with $\mu = {1 \over 2}(1-w_v)$ and $\sigma^2 = {1 \over 4}\sum_{i=1}^{n} w_i^2 - w_v^2$ (see, e.g., \cite{FeixEtAl, SlomZycz}). Drawing upon this idea, we will likewise employ the normal distribution to approximate the distribution of weight for coalitions subject to size constraints.

It is known that a sequences of random variables $M_N$, where $M_N$ is the mean of a sample of size $k$ drawn without replacement from a finite population of size $N$, mean $m$, and variance $s^2$ converges in distribution to $X \sim \cN(m, \sigma)$, where $\sigma = \sqrt{\frac{1}{k} s^2 \left(1 - \frac{k}{N}\right)}$ \cite{ErdosRenyi, Hajek, Rosen, Hoglund}. In our case, we are sampling $k-2$ or $k-1$ countries (depending on whether the coalition is defined to include country $x$) from a population of $N-2$ (inclusion of countries $x$ and $v$ is not random). Accordingly, the distribution of weights for a set $C_{k, \xi}$ of coalitions $c$ such that $\#c = k$ and $\mathbf{1}_{c}(x) = \xi$ can be approximated by the normal distribution with parameters:
\begin{equation} \label{normApproxDM_mu_fixed}
\mu(k, \xi) = (1 - w_v - w_x) \frac{k-1-\xi}{N-2} + w_v + \xi w_x,
\end{equation}
\begin{small}
\begin{equation} \label{normApproxDM_sigma_fixed}
\sigma^2(k, \xi) = \frac{k-1-\xi}{N-2} \left(1 - \frac{k-1-\xi}{N-2}\right) \left(\sum_{i=1}^{N} w_i^2 - w_v^2 - w_x^2 - \left(\frac{1 - w_v - w_x}{N-2}\right)^2 \right),
\end{equation}
\end{small}

while the distribution of weights for a set $C_{k_0, \xi}$ of coalitions $c$ such that $\#c \ge k_0$ and $\mathbf{1}_{c}(x) = \xi$ -- by the normal distribution with parameters:
\begin{align} \label{normApproxDM_mu_int}
\mu_{+}(k_0, \xi) &= \frac{\sum_{k=k_0}^{N-1+\xi} \mu(k, \xi) \binom{N-2}{k-1-\xi}}{\sum_{k=k_0}^{N-1+\xi} \binom{N-2}{k-1-\xi}} = \\
&= (1 - w_v - w_x) \frac{\sum_{k=k_0}^{N-1+\xi} \binom{N-3}{k-2-\xi}}{\sum_{k=k_0}^{N-1+\xi} \binom{N-2}{k-1-\xi}} + w_v + \xi w_x = \\
&= \frac{1 - w_v - w_x}{N-2} \Big(k_0 -1 - \xi + \Theta(N, k_0, \xi)\Big) + w_v + \xi w_x,
\end{align}
where
\begin{equation} \label{normApproxDM_theta}
\Theta(N, k_0, \xi) := \frac{\binom{N-2}{k_0-\xi} \,_2F_1\left({2, k_0 - N - \xi + 2 \over k_0 - \xi + 1}; -1\right)}{\binom{N-2}{k_0-\xi-1} \,_2F_1\left({1, k_0 - N - \xi + 1 \over k_0 - \xi}; -1\right)},
\end{equation}
and
\begin{equation} \label{normApproxDM_sigma_int}
\sigma_{+}^2(k, \xi) = \frac{\sum_{k=k_0}^{N-1+\xi} \Big(\sigma^2(k, \xi) + \mu^2(k, \xi)\Big) \binom{N-2}{k-1-\xi}}{\sum_{k=k_0}^{N-1+\xi} \binom{N-2}{k-1-\xi}} - \mu_{+}^2(k_0, \xi).
\end{equation}
\eqref{normApproxDM_sigma_int} can be expressed in terms of hypergeometric functions as well, but to avoid undue verbosity we will leave in the above form.

\eqref{banChange} can now be expressed in terms of the normal c.d.f. ($\Phi$):
\begin{small}
\begin{align} \label{banChangeNorm}
&\hat{\Delta}(v) = \Phi\Bigg(\frac{\min\Big\{{q_2 (1 - w_x) + w_v \atop q_2}\Big\} - \mu_{+}(K-1, 0)}{\sigma_{+}(K-1, 0)}\Bigg) \sumBinoms{N-2}{K-2} \nonumber \\
& - \Phi\Bigg(\frac{q_2 (1 - w_x) - \mu_{+}(K-1, 0)}{\sigma_{+}(K-1, 0)}\Bigg) \sumBinoms{N-2}{K-2} \nonumber \\
& + \Bigg(1 - \Phi\bigg(\frac{q_2 (1 - w_x) + w_v - \mu(K-1, 0)}{\sigma(K-1, 0)}\bigg)\Bigg) \binom{N-2}{K-2} \nonumber \\
& - \left(\Phi\Bigg(\frac{q_2 + w_v - \mu_{+}(K, 0)}{\sigma_{+}(K, 0)}\Bigg)  + \Phi\Bigg(\frac{\max\Big\{{q_2 (1 - w_x) + w_v \atop q_2}\Big\} - \mu_{+}(K, 0)}{\sigma_{+}(K, 0)}\Bigg)\right) \sumBinoms{N-2}{K-1} \nonumber \\
& - \Bigg(1 + \Phi\bigg(\frac{q_2 + w_v - \mu(K, 0)}{\sigma(K, 0)}\bigg)\Bigg) \binom{N-2}{K-1} \nonumber \\
& - \left(\Phi\Bigg(\frac{\min\Big\{{q_2 + (1 - q_2) w_x \atop q_2 + w_v}\Big\} - \mu_{+}(K, 1)}{\sigma_{+}(K, 1)}\Bigg) + \Phi\Bigg(\frac{q_2 - \mu_{+}(K, 1)}{\sigma_{+}(K, 1)}\Bigg)\right) \sumBinoms{N-2}{K-2} \nonumber \\
& -\max\Bigg(0, \Phi\bigg(\frac{q_2 + (1 - q_2) w_x - \mu(K, 1)}{\sigma(K, 1)}\bigg) - \Phi\bigg(\frac{q_2 + w_v - \mu(K, 1)}{\sigma(K, 1)}\bigg)\Bigg) \binom{N-2}{K-2} \nonumber \\
& + \Phi\Bigg(\frac{q_2 + (1 - q_2) w_x + w_v - \mu_{+}(K+1, 1)}{\sigma_{+}(K+1, 1)}\Bigg) \sumBinoms{N-2}{K-1} \nonumber \\
& - \Phi\Bigg(\frac{\max\Big\{{q_2 + (1 - q_2) w_x \atop q_2 + w_v}\Big\} - \mu_{+}(K+1, 1)}{\sigma_{+}(K+1, 1)}\Bigg)  \sumBinoms{N-2}{K-1},
\end{align}
\end{small}
where
\begin{equation} \label{sumBinomials}
\sumBinoms{n}{k} := \sum_{i=k}^{n} \binom{n}{i} = \binom{n}{k} \,_2F_1\left({1, k-n \over k+1}; -1\right).
\end{equation}

\begin{figure}[h!]
	\centering
	\includegraphics[width=0.92\linewidth]{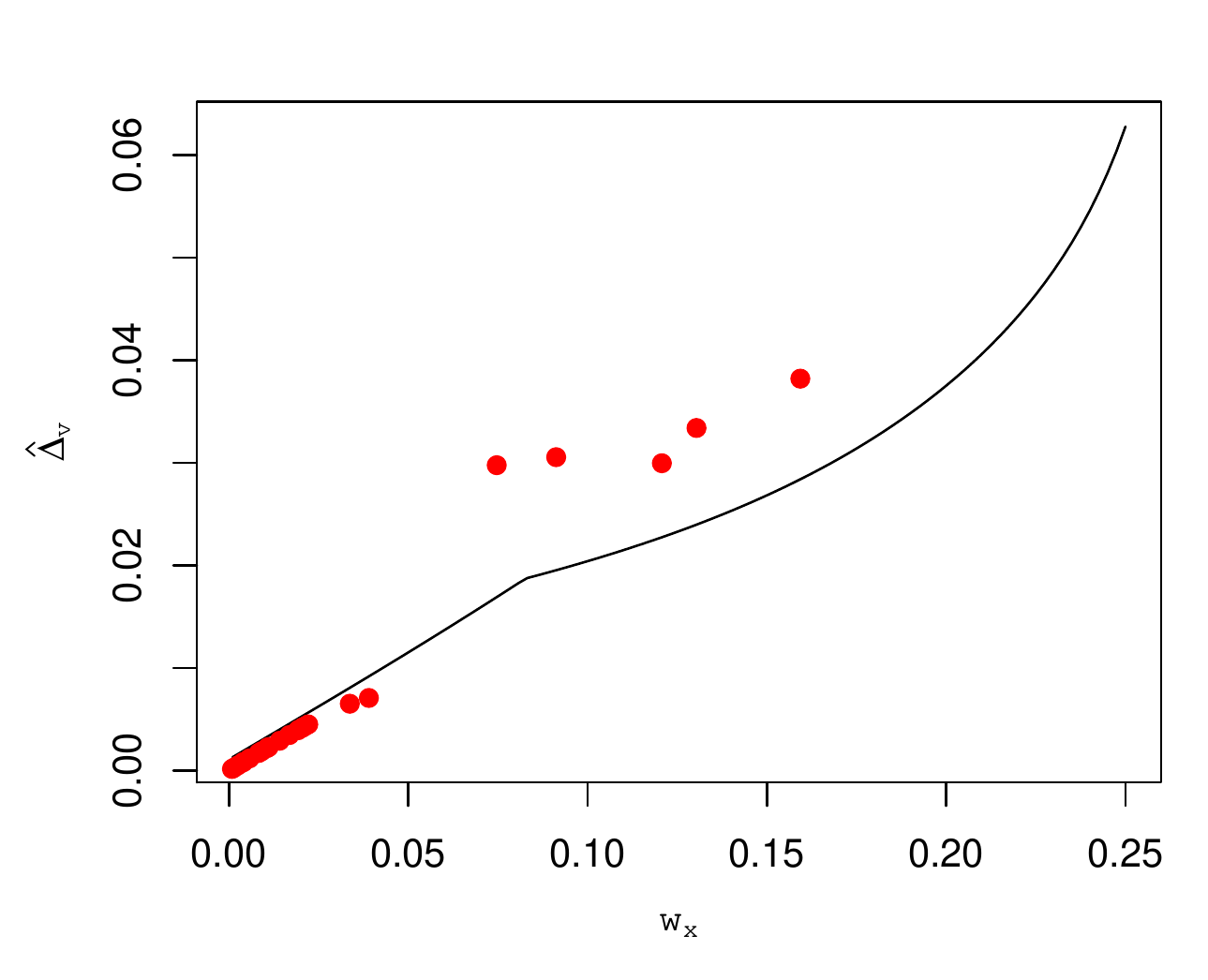}
	\caption{Difference of post-Brexit and pre-Brexit Banzhaf power ($\Delta_v := \phi_v^{27} - \phi_v^{28}$), estimated using normal approximation, as a function of country's voting weight (i.e., normalized population). Red points correspond to exact values of $\Delta_v$ for current EU member states.}
	\label{fig:NormApproxDeltaV}
\end{figure}

\begin{figure}[h!]
	\centering
	\includegraphics[width=0.49\linewidth]{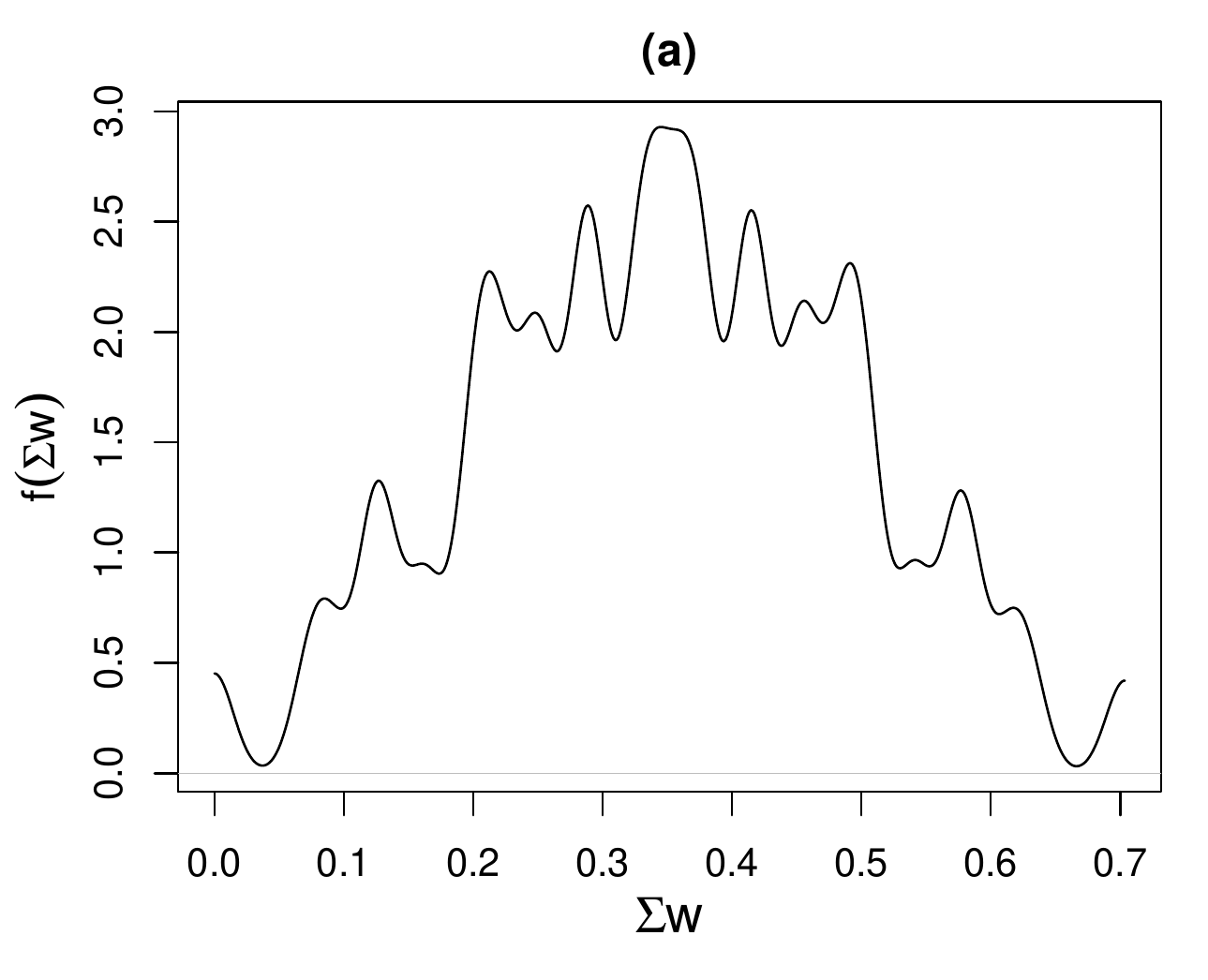}
	\includegraphics[width=0.49\linewidth]{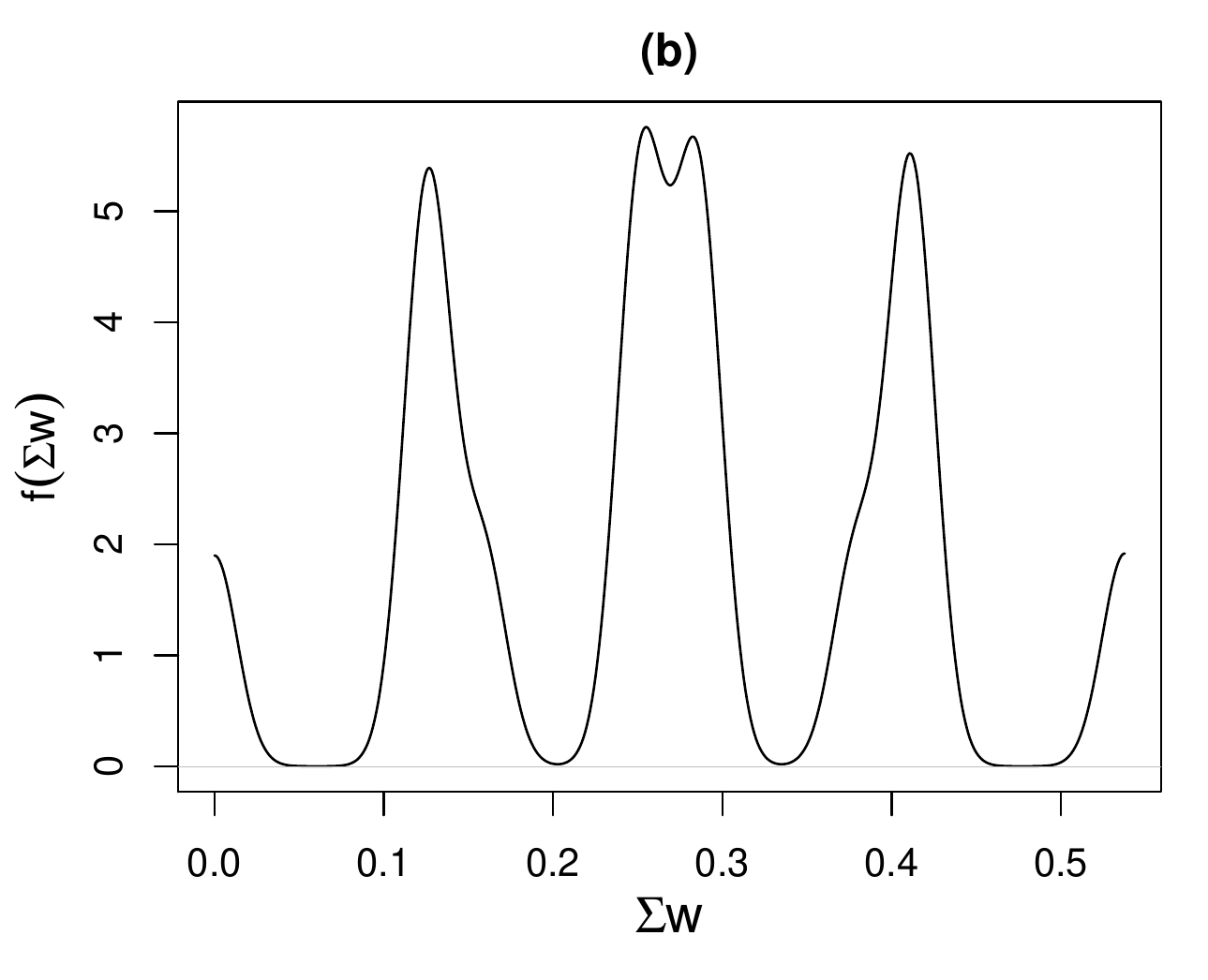}
	\caption{Density functions $f$ of the distributions of coalition weights $\sum w$ for coalitions sampled only from the population of (a) the top 6 EU countries, and (b) the top 4 EU countries (obtained through kernel density estimation). Note how the total variation increases as the size of the population approaches 1.}
	\label{fig:densityWtop64}
\end{figure}

\begin{figure}[h!]
	\centering
	\includegraphics[width=0.49\linewidth]{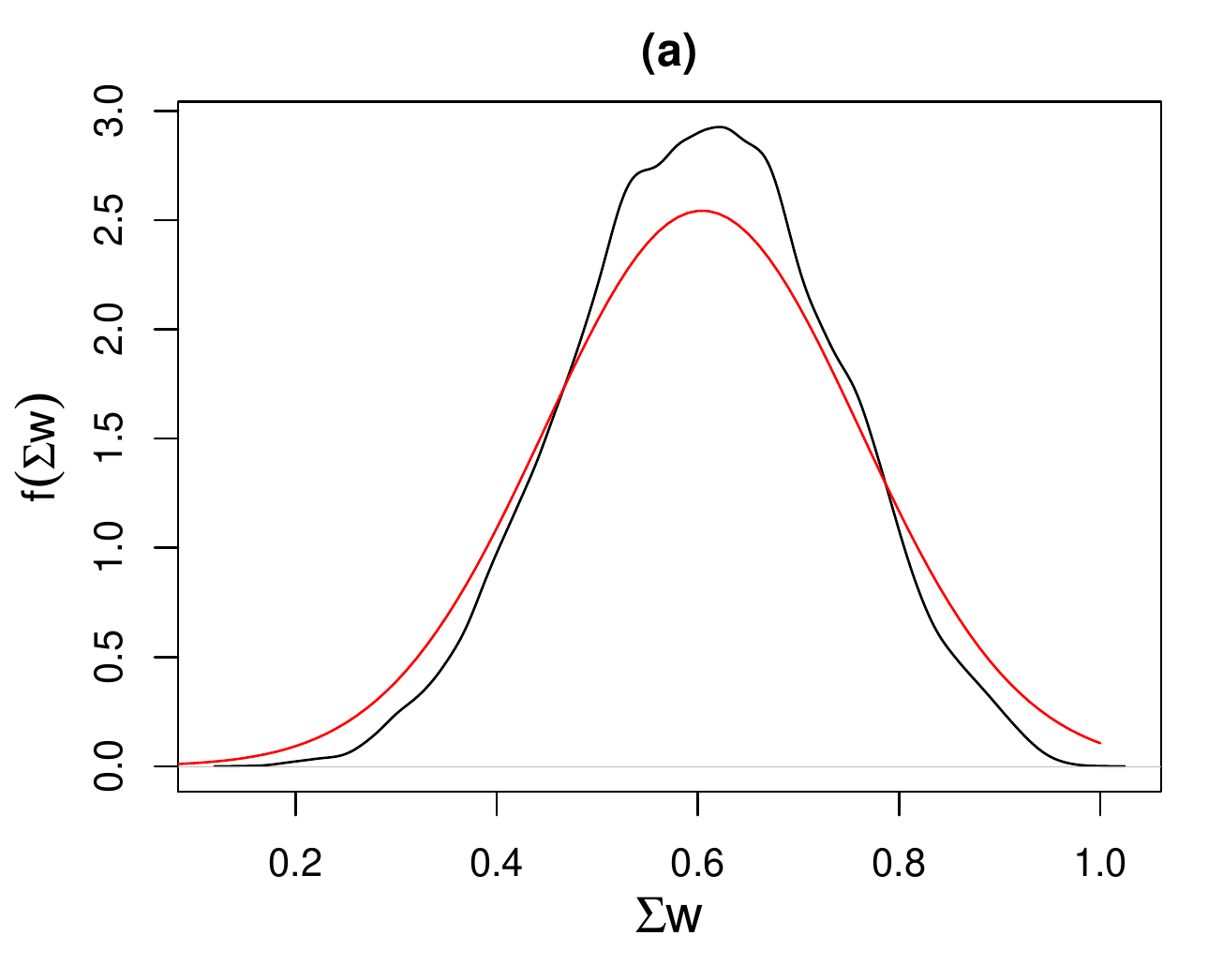}
	\includegraphics[width=0.49\linewidth]{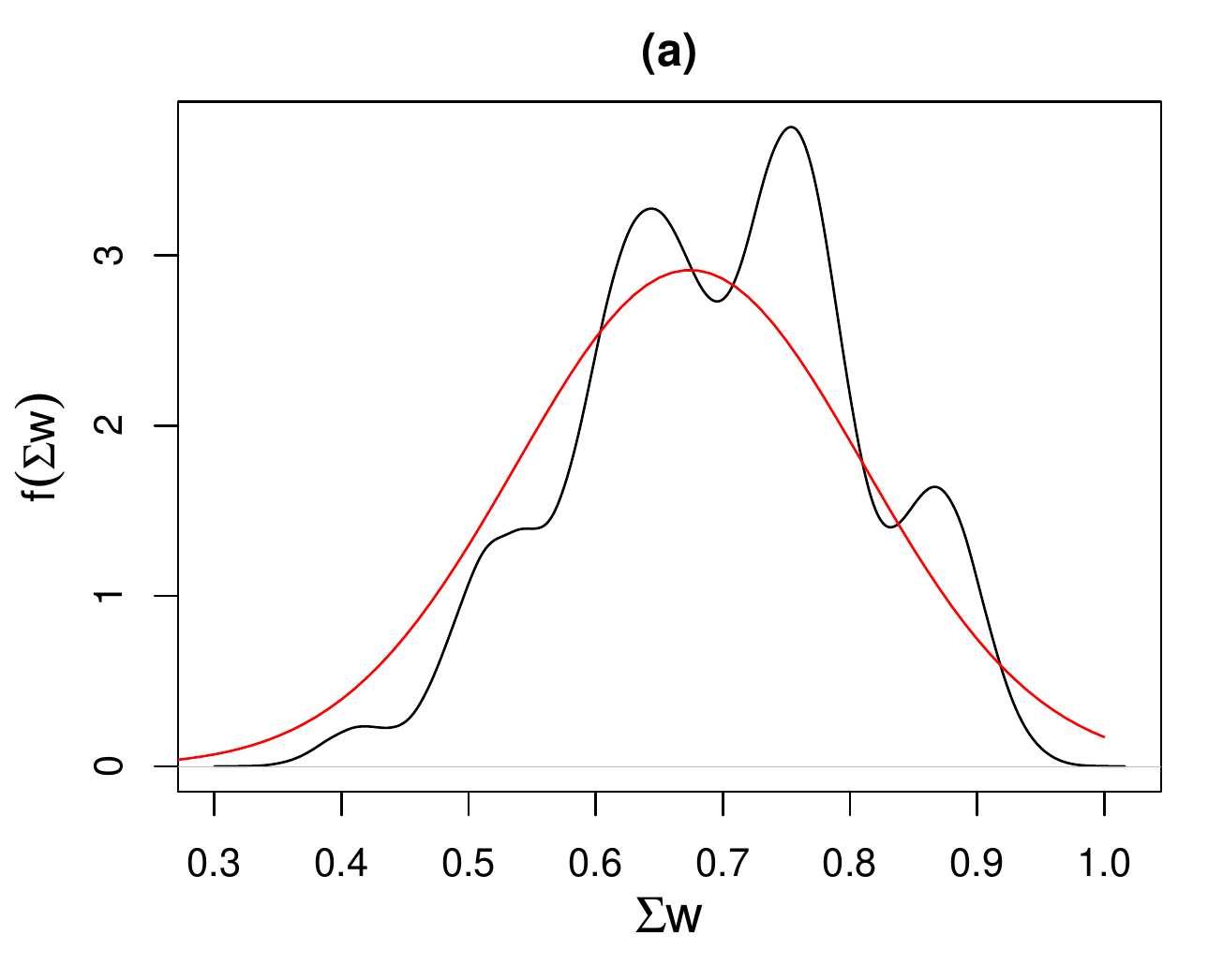}
	\caption{Density functions $f$ of the distributions of coalition weights $\sum w$ for coalitions consisting of (a) Sweden and Denmark, (b) UK and Poland, and 14 countries sampled out of the remaining member states (black line), plotted against the density of the approximating normal distribution with parameters given by \eqref{normApproxDM_mu_fixed} and \eqref{normApproxDM_sigma_fixed} (red line). Quality of the approximation depends on the cardinality of the number of large countries which can be sampled -- six in the first plot, four in the second plot.}
	\label{fig:NormApproxFit}
\end{figure}

As Fig. \ref{fig:NormApproxDeltaV} demonstrates, the normal approximation works rather well for small countries (although there is still a small underestimation), but introduces a significant error for larger countries. The reason has to do with a peculiarity of the EU weight vector: while the large countries account for more than 70\% of the Union's population (70.3604\%, to be exact), there are only six of them. The distribution of voting weights can therefore be thought of in terms of a mixture of Gaussians (each of which approximates quite well the distribution of small countries' weights) centered at several peaks. For six countries, those peaks are numerous enough (there as many as there are subsets of the set of large countries, i.e., $2^6$, although since France, UK, and Italy are very close in terms of population, the number of distinct peaks is actually on the order of $2^5$), as illustrated by Fig. \ref{fig:densityWtop64} (a), that this mixture is approximately unimodal, and can therefore by well approximated by a normal distribution. But when a large country exits the Union \textit{and} we are estimating $\Delta_v$ for another large country, the sampling population of large countries is reduced to 4, the number of distinct peaks decreases exponentially (see Fig. \ref{fig:densityWtop64} (b)), and the overall mixture distribution is no longer approximately normal, as demonstrated by Fig. \ref{fig:NormApproxFit} (b). Its multimodality leads to approximation errors seen on Fig. \ref{fig:NormApproxDeltaV}.

Fig. \ref{fig:NormApproxDeltaV} does not reveal the nonmonotonicities observed on Fig. \ref{fig:brexit}. Those only appear when we divide the change of Banzhaf power by the pre-exit voting power $\psi^{28}(v)$. This quantity can also be estimated using the normal approximation (so we can still describe the effects of an exit by an analytical formula, but at the cost of introducing another source of approximation error):
\begin{align} \label{normApproxBanzhaf}
\psi^{28}(v) &\approx 2^{-28} \binom{N-1}{K-1} \Bigg( 1 - \Phi\Bigg(\frac{q_2 - \mu^{*}(K)}{\sigma^{*}(K)}\Bigg) \Bigg) \nonumber \\
&+ 2^{-28} \binom{N-1}{K} \,_2F_1\left({1, K+1-N \over K}; -1\right) \Phi\Bigg(\frac{q_2 + w_v - \mu_{+}^{*}(K)}{\sigma_{+}^{*}(K)}\Bigg) \nonumber \\
&- 2^{-28} \binom{N-1}{K} \,_2F_1\left({1, K+1-N \over K}; -1\right) \Phi\Bigg(\frac{q_2 - \mu_{+}^{*}(K)}{\sigma_{+}^{*}(K)}\Bigg),
\end{align}
where
\begin{equation}
\mu^{*}(k) = (1 - w_v) \frac{k-1}{N-1} + w_v,
\end{equation}
\begin{equation}
(\sigma^{*})^2(k) = \frac{k-1}{N-1} \left(1 - \frac{k-1}{N-1}\right) \left(\sum_{i=1}^{N} w_i^2 - w_v^2 - \left(\frac{1 - w_v}{N-1}\right)^2 \right),
\end{equation}
\begin{equation}
\mu_{+}^{*}(k_0) = \frac{1 - w_v}{N-1} \Big(k_0 - 1 + \Theta(N, k_0, 0)\Big) + w_v,
\end{equation}
and
\begin{equation}
(\sigma_{+}^{*})^2(k, \xi) = \frac{\sum_{k=k_0}^{N} \Big((\sigma^{*})^2(k) + (\mu^{*})^2(k)\Big) \binom{N-1}{k-1}}{\sum_{k=k_0}^{N} \binom{N-1}{k-1}} - (\mu_{+}^{*})^2(k_0).
\end{equation}

Again, the resulting approximation is better for small countries:

\begin{figure}[h!]
	\centering
	\includegraphics[width=0.92\linewidth]{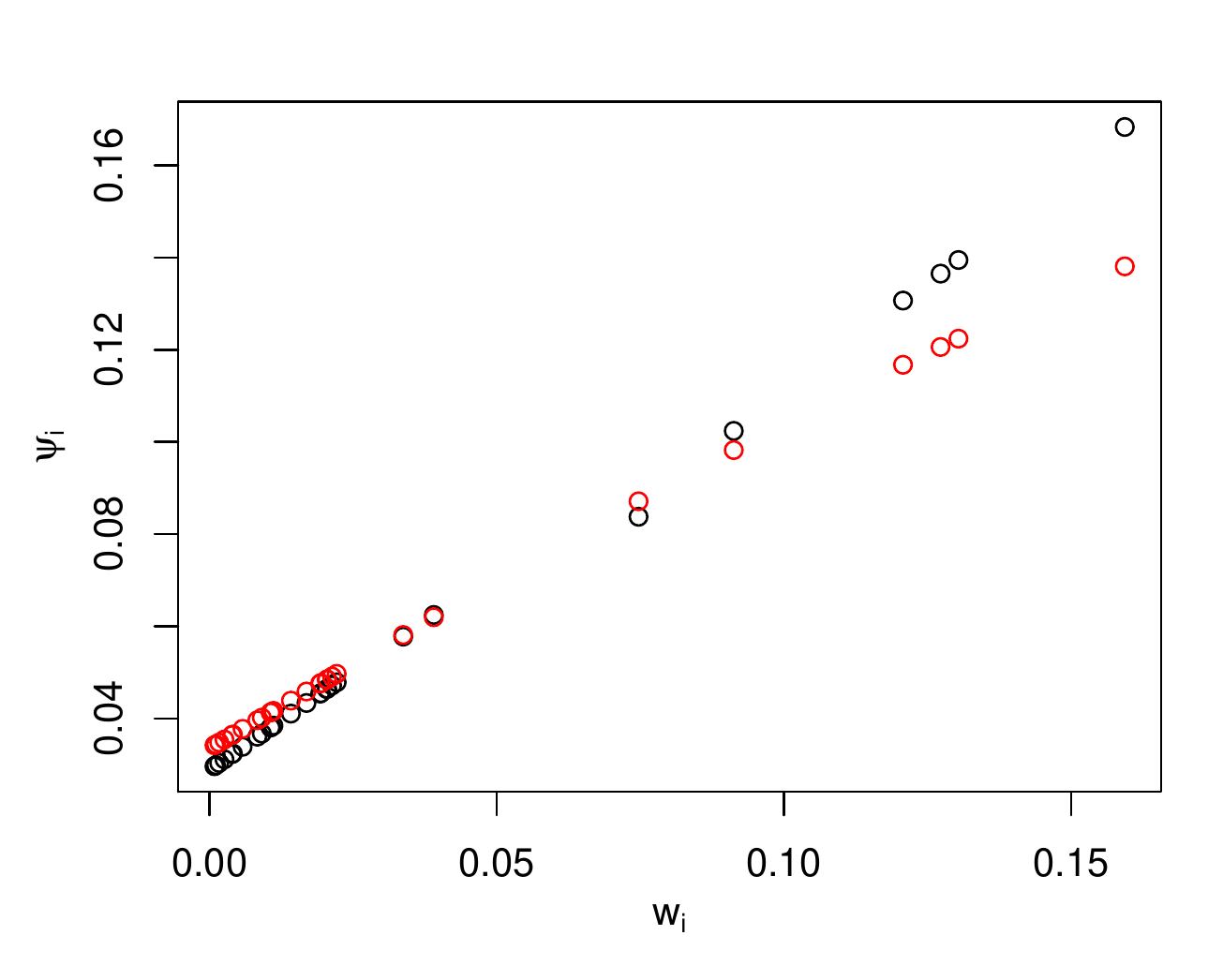}
	\caption{Exact Banzhaf power values of EU member states $\psi(i)$, where $i = 1, ..., 28$ (black points), and their normal approximations (red points) as a function of voting weight $w_i$.}
	\label{fig:banzhafFit}
\end{figure}

\begin{figure}[h!]
	\centering
	\includegraphics[width=0.92\linewidth]{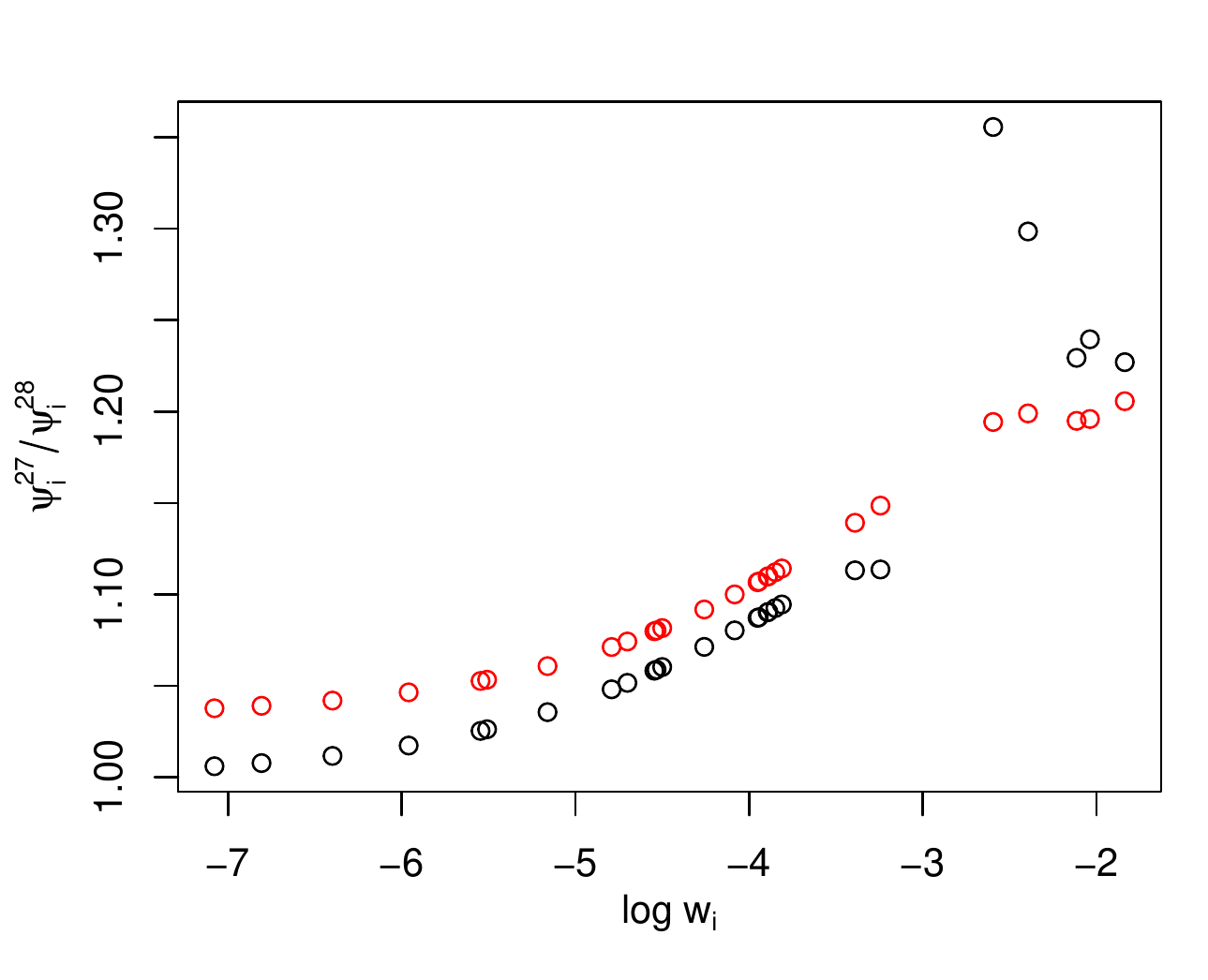}
	\caption{Ratio of post-Brexit and pre-Brexit Banzhaf power values of EU member states other than the UK ($\psi_i^{27}$ and $\psi_i^{27}$, respectively) as a function of voting weight $w_i$. Black points represent exact values, while red points represent normal approximations.}
	\label{fig:banzhafRatio}
\end{figure}

Fig. \ref{fig:banzhafRatio} demonstrates that even under the normal approximation, a different pattern of effects appears for large countries. This suggests that such effects are inherent in the double majority voting rule. Nevertheless, the apparent discontinuity between Romania and Poland and the nonmonotonicity between Poland, Spain, and Italy, appear only for exact values. This in turn indicates that they are caused by the distributional peculiarities of the EU -- small number of large states -- that cause the normal approximation to fail in those cases.

\FloatBarrier

\subsection{N.N.-exit.}

While the difference between Brexit effects for large and small countries has been noted by \cite{Petroczy}, their nonmonotonicity appears to have escaped the attention of earlier researchers. Nor is this effect unique to Brexit: we have analyzed the change of voting power for each of the current 28 members states in the event of every other country leaving the Union of 28 (N.N.-exit). The detailed data are available at our home page, but plots for some representative cases (four large and four small countries) are included below.

\begin{figure}[h]
	\centering
	\includegraphics[width=\linewidth]{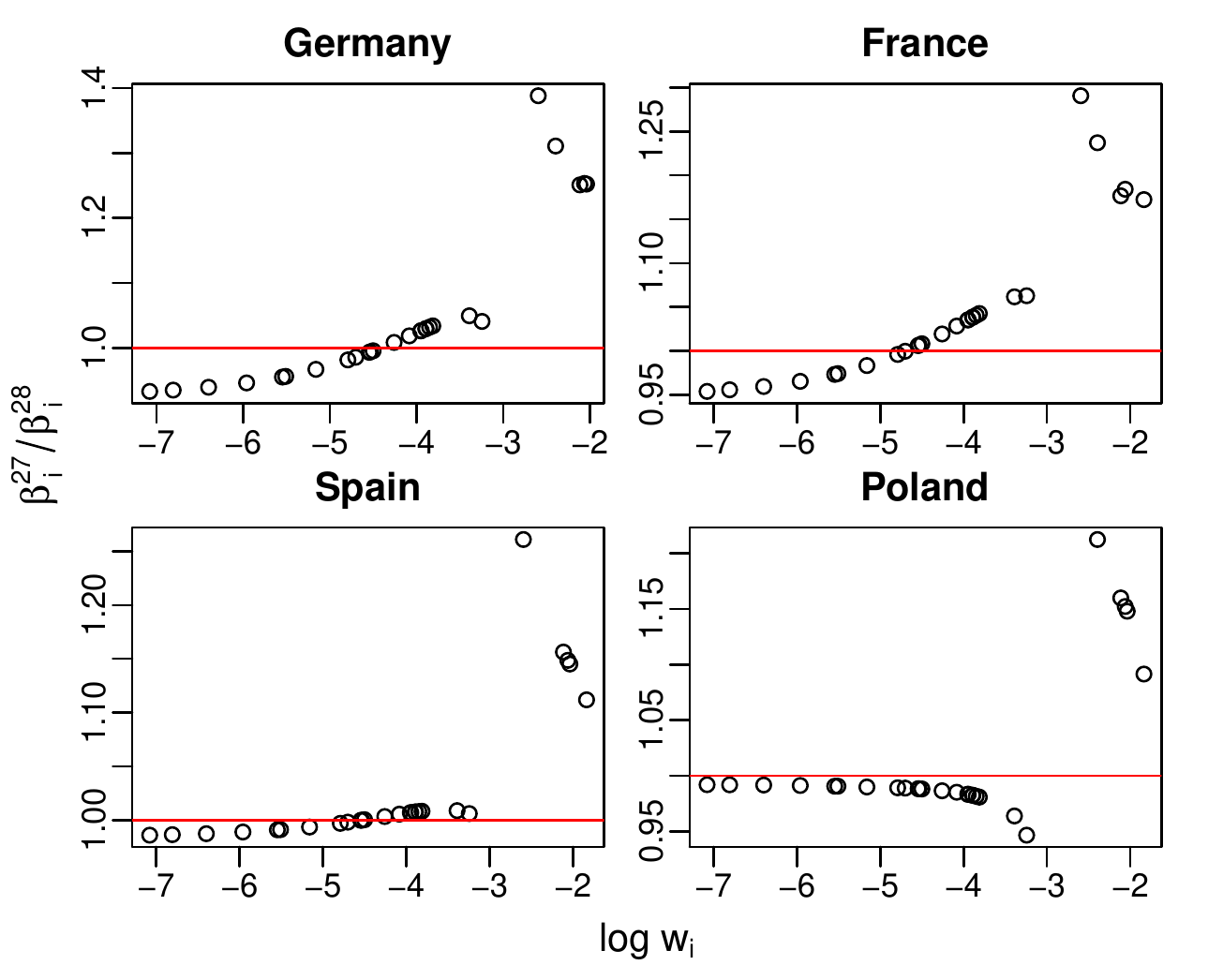}
	\caption{Ratio of post-exit and pre-exit Banzhaf indices of remaining EU member states in the event of a large country leaving the 28-member EU as a function of the pre-exit voting weights.}
	\label{fig:largeCountries}
\end{figure}

\begin{figure}[h]
	\centering
	\includegraphics[width=\linewidth]{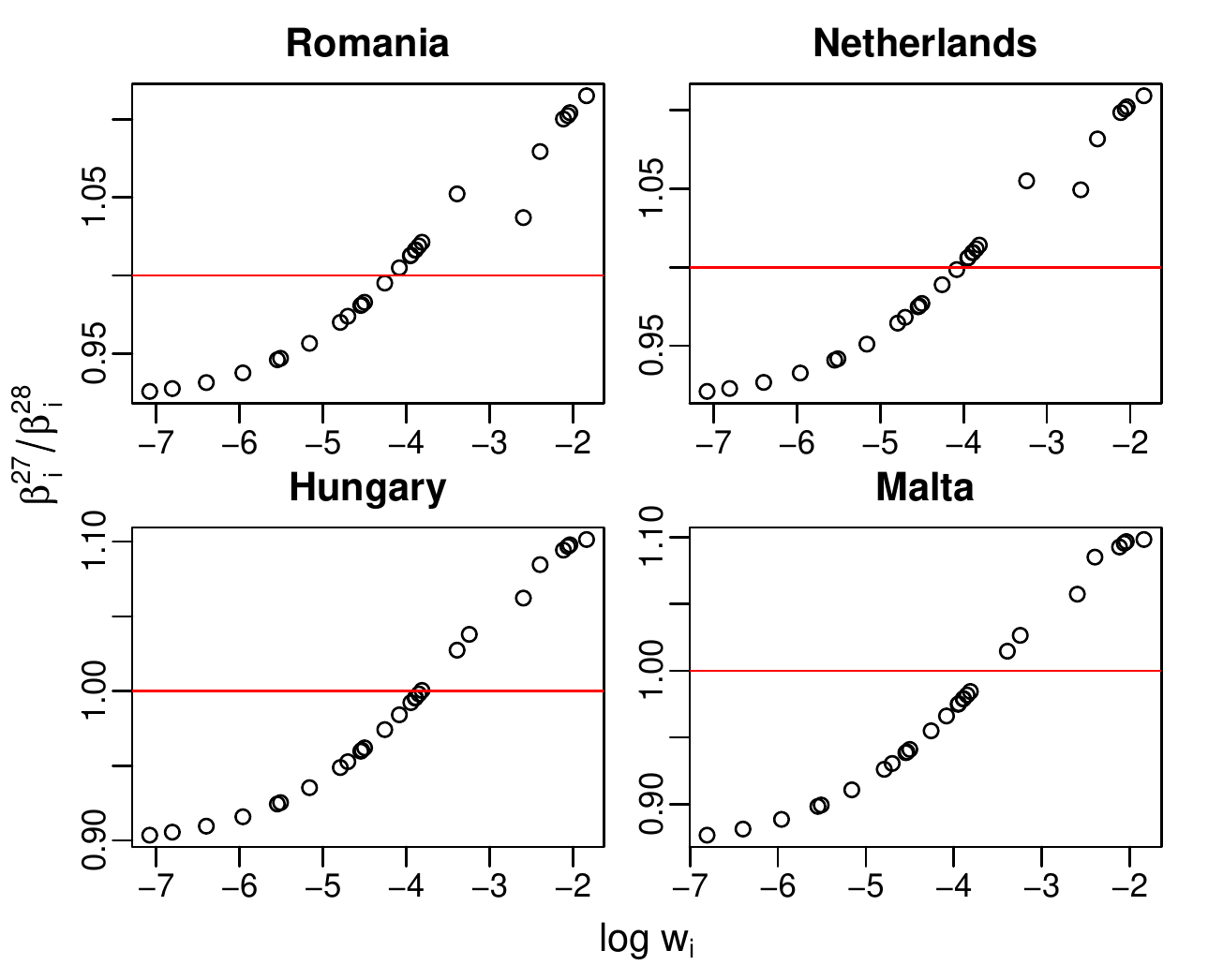}
	\caption{Ratio of post-exit and pre-exit Banzhaf indices of remaining EU member states in the event of a small country leaving the 28-member EU as a function of the pre-exit voting weights.}
	\label{fig:smallCountries}
\end{figure}

Our calculations reveal three patterns of N.N.-exit effects (change of voting power as a function of the original voting weight), with sharp difference between large and small countries:
\begin{itemize}
	\item When a small country leaves the Union, the change of voting power is increasing and convex for small countries, also increasing but concave for large countries, and there appears to be a discontinuity between the two sets of countries;

	\item When a large country other than Poland leaves the Union, the change of voting power is non-monotonic but apparently smooth for small countries (first increasing and convex, later than increasing and concave, and finally decreasing and concave), decreasing for large countries, and a discontinuity exists between the large and small countries, with all values for large countries being above all values for small countries;

	\item When Poland (the smallest large country) leaves the Union, the change of voting power is decreasing and concave for small countries, also decreasing for large countries (with not enough data points to reliably assess convexity), and there is a discontinuity between the two sets of countries, with all values for large countries being above all values for small countries.
\end{itemize}

We conjecture that those patterns have not been noted with earlier researchers, as they have preoccupied primarily with the scenario of another member state leaving the EU of 27 (post-Brexit). This would be a very different case, as it would involve no change in the absolute threshold under the first voting rule, since $\lceil q_1 27 \rceil = 15 = \lceil q_1 26 \rceil$. But if we assume that two countries leave the current EU of 27, and analyze the exit of a potential third country, the patterns discussed above reappear.

\begin{figure}[h]
	\centering
	\includegraphics[width=\linewidth]{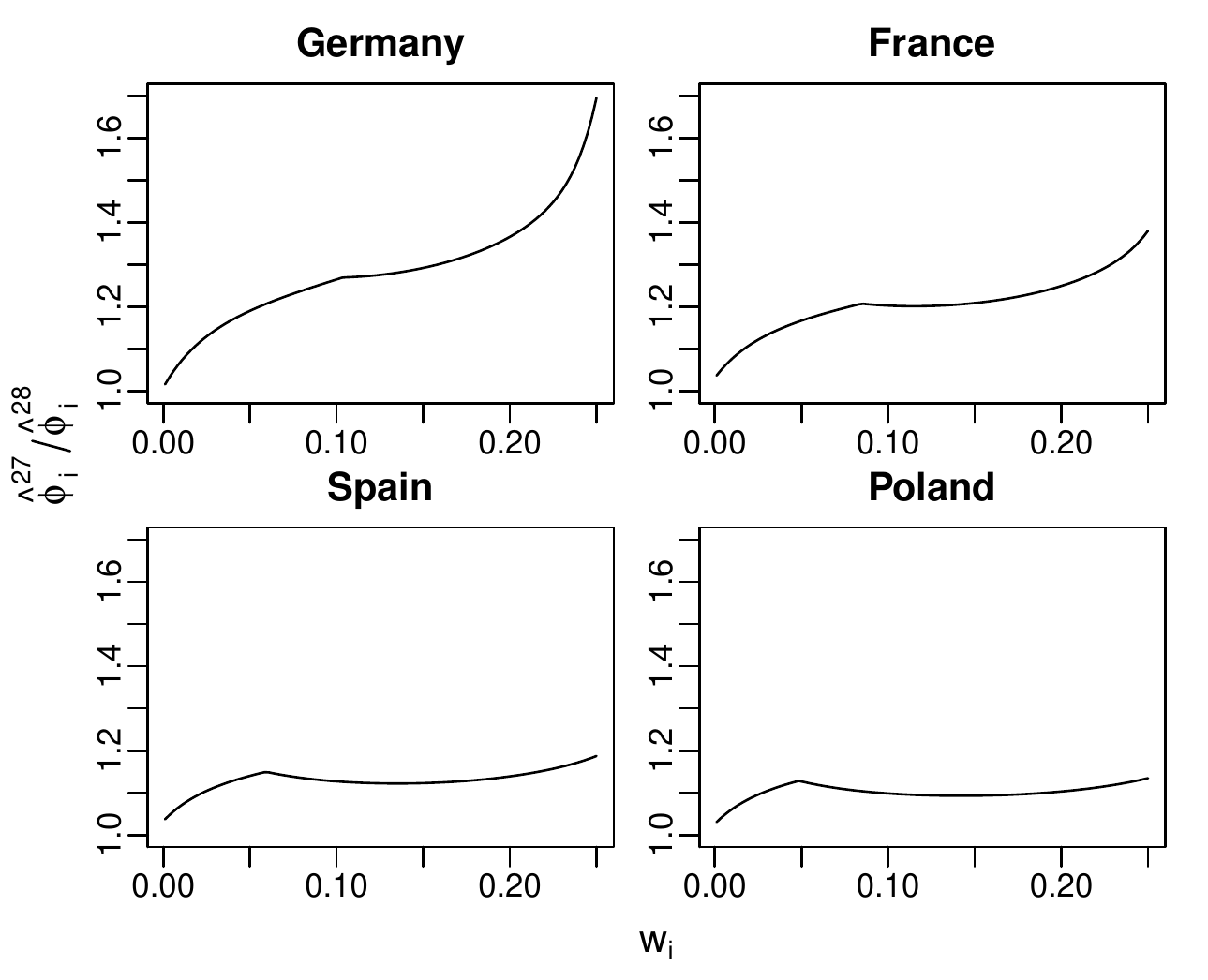}
	\caption{Estimated ratio of post-exit and pre-exit Banzhaf power values of remaining EU member states in the event of a large country leaving the 28-member EU, obtained through normal approximation of weight distributions, as a function of the pre-exit voting weights.}
	\label{fig:largeApprox}
\end{figure}

\begin{figure}[h]
	\centering
	\includegraphics[width=\linewidth]{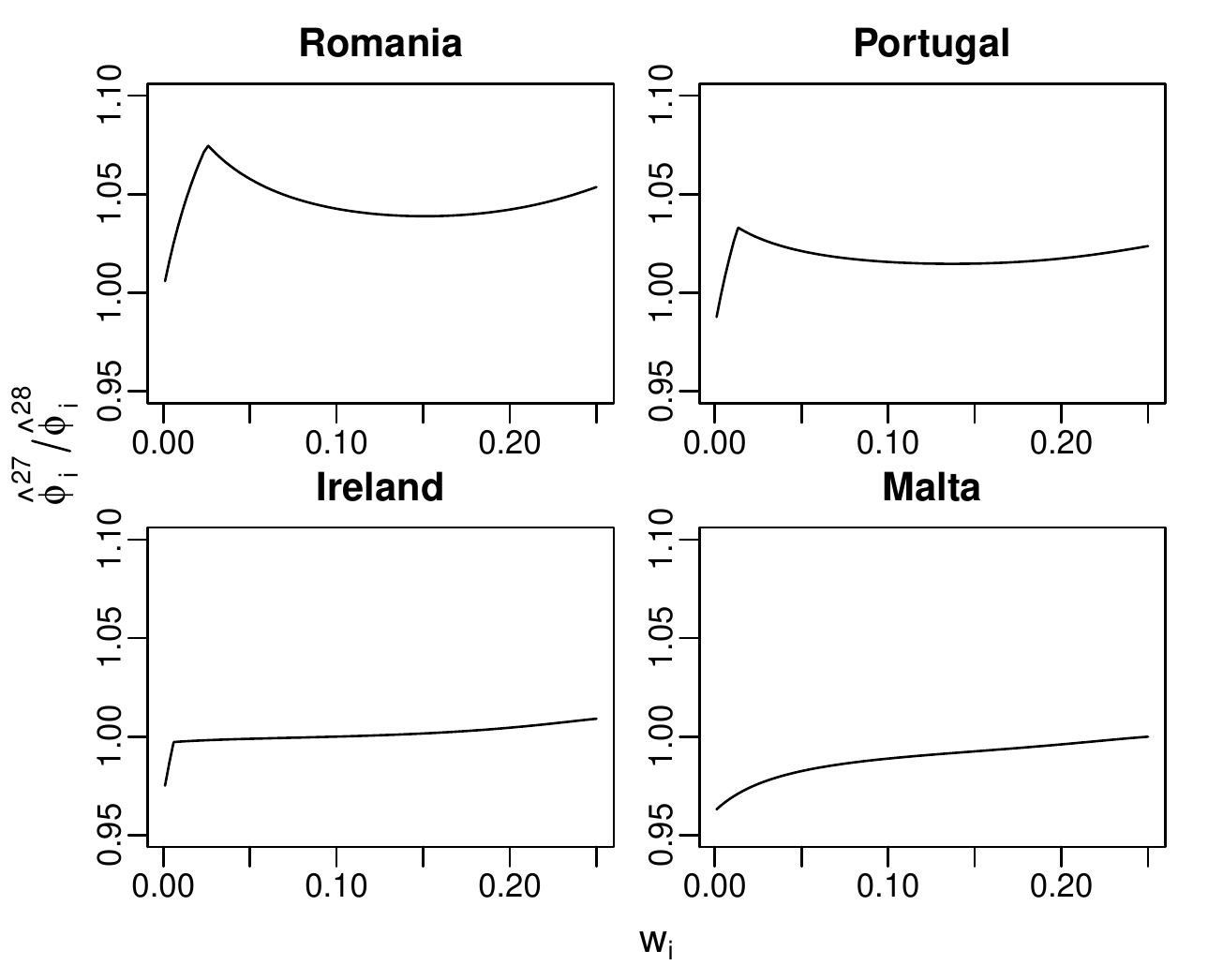}
	\caption{Estimated ratio of post-exit and pre-exit Banzhaf power values of remaining EU member states in the event of a small country leaving the 28-member EU, obtained through normal approximation of weight distributions, as a function of the pre-exit voting weights.}
	\label{fig:smallApprox}
\end{figure}

At least in part those different patterns can be explained by reference to the approximation method discussed in the foregoing section. Figs. \ref{fig:largeApprox} and \ref{fig:smallApprox} illustrate how the ratio of post-exit and pre-exit Banzhaf power values as a function of pre-exit voting weights would change depending on the size of the leaving country.

\bigskip\bigskip
\newpage
\section{Appendix A}
\begin{center}
\begin{longtable}{|l|c|c|c|c|}\hline
Country & Population & \multicolumn{2}{c|}{Banzhaf Index (\%)}   & Relative\\
& (millions)& with UK & without UK&Difference\\
\hline &&&&\\[-4.5mm]\hline
Germany	&	81.09	&	15.89 \%	&	18.21 \%	&	14.61 \%	\\ \hline
France	&	66.35	&	13.14 \%	&	15.12 \%	&	15.05 \%	\\ \hline
United Kingdom	&	64.77	&	12.83 \%	&		&		\\ \hline
Italy	&	61.44	&	12.15 \%	&	13.91 \%	&	14.48 \%	\\ \hline
Spain	&	46.44	&	9.17 \%	&	10.66 \%	&	16.29 \%	\\ \hline
Poland	&	38.01	&	7.01 \%	&	9.15 \%	&	30.63 \%	\\ \hline
Romania	&	19.86	&	3.94 \%	&	4.30 \%	&	9.18 \%	\\ \hline
Netherlands	&	17.16	&	3.39 \%	&	3.73 \%	&	9.92 \%	\\ \hline
Belgium	&	11.26	&	2.22 \%	&	2.46 \%	&	10.66 \%	\\ \hline
Greece	&	10.85	&	2.14 \%	&	2.37 \%	&	10.69 \%	\\ \hline
Czech Republic	&	10.42	&	2.06 \%	&	2.28 \%	&	10.71 \%	\\ \hline
Portugal	&	10.37	&	2.05 \%	&	2.27 \%	&	10.71 \%	\\ \hline
Hungary	&	9.86	&	1.95 \%	&	2.16 \%	&	10.75 \%	\\ \hline
Sweden	&	9.79	&	1.93 \%	&	2.14 \%	&	10.75 \%	\\ \hline
Austria	&	8.58	&	1.70 \%	&	1.88 \%	&	10.83 \%	\\ \hline
Bulgaria	&	7.20	&	1.43 \%	&	1.58 \%	&	10.86 \%	\\ \hline
Denmark	&	5.65	&	1.11 \%	&	1.24 \%	&	10.94 \%	\\ \hline
Finland	&	5.47	&	1.08 \%	&	1.20 \%	&	10.94 \%	\\ \hline
Slovakia	&	5.40	&	1.06 \%	&	1.18 \%	&	10.93 \%	\\ \hline
Ireland	&	4.63	&	0.91 \%	&	1.01 \%	&	10.95 \%	\\ \hline
Croatia	&	4.23	&	0.83 \%	&	0.92 \%	&	10.96 \%	\\ \hline
Lithuania	&	2.92	&	0.57 \%	&	0.64 \%	&	11.07 \%	\\ \hline
Slovenia	&	2.06	&	0.41 \%	&	0.46 \%	&	10.94 \%	\\ \hline
Latvia	&	1.99	&	0.39 \%	&	0.43 \%	&	10.94 \%	\\ \hline
Estonia	&	1.31	&	0.26 \%	&	0.29 \%	&	10.88 \%	\\ \hline
Cyprus	&	0.85	&	0.17 \%	&	0.19 \%	&	11.00 \%	\\ \hline
Luxembourg	&	0.56	&	0.11 \%	&	0.12 \%	&	10.92 \%	\\ \hline
Malta	&	0.43	&	0.08 \%	&	0.09 \%	&	10.84 \%	\\ \hline

\caption{Banzhaf Indices for $\cV^{1}$ before and after Brexit}\label{tab1}
\end{longtable}
\end{center}
\begin{table}
\begin{center}
\begin{tabular}{|l|c|c|c|c|c|}\hline
Country & Population & \multicolumn{2}{c|}{Banzhaf Index (\%)}  & Relative\\
& (millions)& with UK & without UK&Difference\\
\hline &&&&\\[-4.5mm]\hline
Germany	&	81.09	&	10.19\% 	&	11.89\% 	&	16.69\% 	\\ \hline
France	&	66.35	&	8.45\% 	&	9.96\% 	&	17.89\% 	\\ \hline
United Kingdom	&	64.77	&	8.27\% 	&		&		\\ \hline
Italy	&	61.44	&	7.91\% 	&	9.25\% 	&	16.92\% 	\\ \hline
Spain	&	46.44	&	6.20\% 	&	7.65\% 	&	23.52\% 	\\ \hline
Poland	&	38.01	&	5.07\% 	&	6.54\% 	&	28.87\% 	\\ \hline
Romania	&	19.86	&	3.78\% 	&	4.00\% 	&	5.90\% 	\\ \hline
Netherlands	&	17.16	&	3.50\% 	&	3.70\% 	&	5.85\% 	\\ \hline
Belgium	&	11.26	&	2.90\% 	&	3.01\% 	&	4.07\% 	\\ \hline
Greece	&	10.85	&	2.86\% 	&	2.97\% 	&	3.88\% 	\\ \hline
Czech Republic	&	10.42	&	2.81\% 	&	2.92\% 	&	3.69\% 	\\ \hline
Portugal	&	10.37	&	2.81\% 	&	2.91\% 	&	3.67\% 	\\ \hline
Hungary	&	9.86&	2.76\% 	&	2.85\% 	&	3.41\% 	\\ \hline
Sweden	&	9.79	&	2.75\% 	&	2.84\% 	&	3.36\% 	\\ \hline
Austria	&	8.58	&	2.63\% 	&	2.70\% 	&	2.73\% 	\\ \hline
Bulgaria	&	7.20	&	2.49\% 	&	2.54\% 	&	1.88\% 	\\ \hline
Denmark	&	5.65	&	2.33\% 	&	2.35\% 	&	0.81\% 	\\ \hline
Finland	&	5.47	&	2.31\% 	&	2.33\% 	&	0.69\% 	\\ \hline
Slovakia	&	5.40	&	2.30\% 	&	2.32\% 	&	0.61\% 	\\ \hline
Ireland	&	4.63	&	2.22\% 	&	2.22\% 	&	0.00\% 	\\ \hline
Croatia	&	4.23	&	2.18\% 	&	2.18\% 	&	-0.34\% 	\\ \hline
Lithuania	&	2.92	&	2.05\% 	&	2.02\% 	&	-1.56\% 	\\ \hline
Slovenia	&	2.06	&	1.96\% 	&	1.92\% 	&	-2.40\% 	\\ \hline
Latvia	&	1.99	&	1.95\% 	&	1.91\% 	&	-2.51\% 	\\ \hline
Estonia	&	1.31	&	1.89\% 	&	1.82\% 	&	-3.26\% 	\\ \hline
Cyprus	&	0.85	&	1.84\% 	&	1.77\% 	&	-3.80\% 	\\ \hline
Luxembourg	&	0.56	&	1.81\% 	&	1.73\% 	&	-4.19\% 	\\ \hline
Malta	&	0.43	&	1.79\% 	&	1.71\% 	&	-4.38\% 	\\ \hline

\end{tabular}\\
\caption{Banzhaf Indices before and after Brexit for the Lisbon system}\label{tab2}\end{center}
\end{table}

\begin{table}
\begin{center}
\begin{tabular}{|l|c|c|c|c|c|}\hline
Country & Population & \multicolumn{2}{c|}{Shapley-Shubik Index (\%)}  & Relative\\
& (millions)& with UK & without UK&Difference\\
\hline &&&&\\[-4.5mm]\hline
Germany	&	81.09	&	14.38 \%	&	17.27 \%	&	20.09 \%	\\ \hline
France	&	66.35	&	11.22 \%	&	13.26 \%	&	18.15 \%	\\ \hline
United Kingdom	&	64.77	&	10.91 \%	&		&		\\ \hline
Italy	&	61.44	&	10.27 \%	&	12.15 \%	&	18.33 \%	\\ \hline
Spain	&	46.44	&	7.51 \%	&	8.99 \%	&	19.69 \%	\\ \hline
Poland	&	38.01	&	6.32 \%	&	6.98 \%	&	10.46 \%	\\ \hline
Romania	&	19.86	&	3.74 \%	&	3.98 \%	&	6.56 \%	\\ \hline
Netherlands	&	17.16	&	3.31 \%	&	3.55 \%	&	7.12 \%	\\ \hline
Belgium	&	11.26	&	2.42 \%	&	2.59 \%	&	7.27 \%	\\ \hline
Greece	&	10.85	&	2.36 \%	&	2.52 \%	&	7.13 \%	\\ \hline
Czech Republic	&	10.42	&	2.30 \%	&	2.46 \%	&	7.04 \%	\\ \hline
Portugal	&	10.37	&	2.29 \%	&	2.45 \%	&	7.07 \%	\\ \hline
Hungary	&	9.86	&	2.21 \%	&	2.37 \%	&	6.96 \%	\\ \hline
Sweden	&	9.79	&	2.20 \%	&	2.35 \%	&	6.99 \%	\\ \hline
Austria	&	8.58	&	2.03 \%	&	2.17 \%	&	6.83 \%	\\ \hline
Bulgaria	&	7.20	&	1.83 \%	&	1.94 \%	&	6.08 \%	\\ \hline
Denmark	&	5.65	&	1.61 \%	&	1.68 \%	&	4.60 \%	\\ \hline
Finland	&	5.47	&	1.58 \%	&	1.66 \%	&	4.54 \%	\\ \hline
Slovakia	&	5.40	&	1.57 \%	&	1.64 \%	&	4.47 \%	\\ \hline
Ireland	&	4.63	&	1.46 \%	&	1.51 \%	&	3.61 \%	\\ \hline
Croatia	&	4.23	&	1.41 \%	&	1.45 \%	&	3.14 \%	\\ \hline
Lithuania	&	2.92	&	1.22 \%	&	1.24 \%	&	2.21 \%	\\ \hline
Slovenia	&	2.06	&	1.10 \%	&	1.10 \%	&	0.36 \%	\\ \hline
Latvia	&	1.99	&	1.09 \%	&	1.09 \%	&	0.13 \%	\\ \hline
Estonia	&	1.31	&	0.99 \%	&	0.98 \%	&	-1.22 \%	\\ \hline
Cyprus	&	0.85	&	0.93 \%	&	0.91 \%	&	-2.11 \%	\\ \hline
Luxembourg	&	0.56	&	0.89 \%	&	0.86 \%	&	-2.97 \%	\\ \hline
Malta	&	0.43	&	0.87 \%	&	0.84 \%	&	-3.50 \%	\\ \hline
\end{tabular}\\
\caption{Shapley-Shubik Indices before and after Brexit for the Lisbon system}\label{tab2S}\end{center}
\end{table}
\begin{table}
\begin{center}
\begin{tabular}{|l|c|c|c|c|c|}\hline
Country & Population & \multicolumn{2}{c|}{Banzhaf Index (\%)}  & Relative\\
& (millions)& with UK & without UK&Difference\\
\hline &&&&\\[-4.5mm]\hline
Germany	&	81.09	&	10.19 \% 	&	9.92 \% 	&	-2.67 \% 	\\ \hline
France	&	66.35	&	8.45 \% 	&	8.39 \% 	&	-0.61 \% 	\\ \hline
United Kingdom	&	64.77	&	8.27 \% 	&		&		\\ \hline
Italy	&	61.44	&	7.91 \% 	&	7.84 \% 	&	-0.96 \% 	\\ \hline
Spain	&	46.44	&	6.20 \% 	&	6.67 \% 	&	7.70 \% 	\\ \hline
Poland	&	38.01	&	5.07 \% 	&	5.66 \% 	&	11.46 \% 	\\ \hline
Romania	&	19.86	&	3.78 \% 	&	3.91 \% 	&	3.52 \% 	\\ \hline
Netherlands	&	17.16	&	3.50 \% 	&	3.69 \% 	&	5.60 \% 	\\ \hline
Belgium	&	11.26	&	2.90 \% 	&	3.18 \% 	&	9.90 \% 	\\ \hline
Greece	&	10.85	&	2.86 \% 	&	3.15 \% 	&	10.25 \% 	\\ \hline
Czech Republic	&	10.42	&	2.81 \% 	&	3.11 \% 	&	10.61 \% 	\\ \hline
Portugal	&	10.37	&	2.81 \% 	&	3.11 \% 	&	10.66 \% 	\\ \hline
Hungary	&	9.86 &	2.76 \% 	&	3.06 \% 	&	11.11 \% 	\\ \hline
Sweden	&	9.79	&	2.75 \% 	&	3.05 \% 	&	11.21 \% 	\\ \hline
Austria	&	8.58	&	2.63 \% 	&	2.95 \% 	&	12.35 \% 	\\ \hline
Bulgaria	&	7.20	&	2.49 \% 	&	2.83 \% 	&	13.79 \% 	\\ \hline
Denmark	&	5.65	&	2.33 \% 	&	2.69 \% 	&	15.69 \% 	\\ \hline
Finland	&	5.47	&	2.31 \% 	&	2.68 \% 	&	15.88 \% 	\\ \hline
Slovakia	&	5.40	&	2.30 \% 	&	2.67 \% 	&	16.01 \% 	\\ \hline
Ireland	&	4.63	&	2.22 \% 	&	2.60 \% 	&	17.03 \% 	\\ \hline
Croatia	&	4.23	&	2.18 \% 	&	2.57 \% 	&	17.60 \% 	\\ \hline
Lithuania	&	2.92	&	2.05 \% 	&	2.45 \% 	&	19.65 \% 	\\ \hline
Slovenia	&	2.06	&	1.96 \% 	&	2.38 \% 	&	21.06 \% 	\\ \hline
Latvia	&	1.99	&	1.95 \% 	&	2.37 \% 	&	21.24 \% 	\\ \hline
Estonia	&	1.31	&	1.89 \% 	&	2.31 \% 	&	22.48 \% 	\\ \hline
Cyprus	&	0.85	&	1.84 \% 	&	2.27 \% 	&	23.40 \% 	\\ \hline
Luxembourg	&	0.56	&	1.81 \% 	&	2.24 \% 	&	24.03 \% 	\\ \hline
Malta	&	0.43	&	1.79 \% 	&	2.23 \% 	&	24.36 \% 	\\ \hline
\end{tabular}\\
\caption{Indices for a modified Lisbon system with quota $q=16$ in $\cV^{2}_{27}$}\label{tab3}\end{center}
\end{table}
\begin{table}
\begin{center}
\begin{tabular}{|l|c|c|c|c|}\hline
Country & Population & \multicolumn{2}{c|}{Banzhaf Index (\%)}  & Relative\\
& (millions)& with UK & without UK&Difference\\
\hline &&&&\\[-4.5mm]\hline
Germany&	81.09&		9.10 \%&	9.89 \%&	8.75 \%\\ \hline
France&	66.35&		8.24 \%&	8.97 \%&	8.87 \%\\ \hline
United Kingdom&	64.77	&	8.14 \%&	 & \\ \hline	
Italy&	61.44	&	7.93 \%	&8.64 \%&	8.89 \%\\ \hline
Spain&	46.44&		6.90 \%&	7.52 \%&	8.91 \%\\ \hline
Poland&	38.01&		6.24 \%&	6.80 \%&	8.92 \%\\ \hline
Romania&	19.86&		4.51 \%&	4.91 \%&	8.88 \%\\ \hline
Netherlands&	17.16&		4.19 \%&	4.56 \%	&8.88 \%\\ \hline
Belgium&	11.26&		3.39 \%&	3.69 \%	&8.87 \%\\ \hline
Greece&	10.855		&3.33 \%	&3.63 \%&	8.87 \%\\ \hline
Czech Republic&	10.42&		3.26 \%&	3.55 \%	&8.87 \%\\ \hline
Portugal	&10.37	&	3.26 \%&	3.55 \%&	8.87 \%\\ \hline
Hungary	&9.86&	3.17 \%  &	3.46 \%  &	8.87 \%  \\ \hline
Sweden	&9.79 &	3.16 \%  &	3.44 \%  &	8.87 \%  \\ \hline
Austria	&8.58	&	2.96 \%  &	3.22 \%  &	8.87 \%  \\ \hline
Bulgaria&	7.20	&	2.71 \%  &	2.95 \%  &	8.88 \%  \\ \hline
Denmark&	5.65	&	2.40 \%  &	2.62 \%  &	8.86 \%  \\ \hline
Finland&	5.47	&	2.36 \%  &	2.57 \%  &	8.85 \%  \\ \hline
Slovakia&	5.40	&	2.35 \%  &	2.56 \%  &	8.86 \%  \\ \hline
Ireland&	4.63&		2.17 \%  &	2.37 \%  &	8.86 \%  \\ \hline
Croatia&	4.23	&	2.08 \%  &	2.26 \%  &	8.86 \%  \\ \hline
Lithuania&	2.92	&	1.73 \%  &	1.88 \%  &	8.87 \%  \\ \hline
Slovenia&	2.06&		1.45 \%  &	1.58 \%  &	8.86 \%  \\ \hline
Latvia&	1.99&		1.42 \%  &	1.55 \%  &	8.86 \%  \\ \hline
Estonia	&1.31	&	1.16 \%  &	1.26 \%  &	8.85 \%  \\ \hline
Cyprus&	0.85	&	0.93 \%  &	1.01 \%  &	8.86 \%  \\ \hline
Luxembourg&	0.56	&	0.76 \%  &	0.82 \%  &	8.85 \%  \\ \hline
Malta &	0.43	&	0.66 \%  &	0.72 \%  &	8.88 \%  \\ \hline
\end{tabular}\\
\caption{Indices for the Jagiellonian Compromise}\label{tab4}\end{center}
\end{table}

\begin{table}
\begin{center}
\begin{tabular}{|l|c|c|c|c|}\hline
Country & Population & \multicolumn{2}{c|}{Banzhaf Index (\%)}  & Relative\\
& (millions)& with UK & without UK&Difference\\
\hline &&&&\\[-4.5mm]\hline
Germany	&	81.09	&	11.89 \%	&	10.61 \%	&	-10.78 \%	\\ \hline
France	&	66.35	&	9.96 \%	&	8.89 \%	&	-10.67 \%	\\ \hline
Scotland	&	5.34	&		&	2.45 \%	&		\\ \hline
Italy	&	61.44	&	9.25 \%	&	8.27 \%	&	-10.59 \%	\\ \hline
Spain	&	46.44	&	7.65 \%	&	6.93 \%	&	-9.47 \%	\\ \hline
Poland	&	38.01	&	6.54 \%	&	5.78 \%	&	-11.59 \%	\\ \hline
Romania	&	19.86	&	4.00 \%	&	3.87 \%	&	-3.36 \%	\\ \hline
Netherlands	&	17.16	&	3.70 \%	&	3.62 \%	&	-2.34 \%	\\ \hline
Belgium	&	11.26	&	3.01 \%	&	3.03 \%	&	0.65 \%	\\ \hline
Greece	&	10.85	&	2.97 \%	&	2.99 \%	&	0.92 \%	\\ \hline
Czech Republic	&	10.42	&	2.92 \%	&	2.95 \%	&	1.22 \%	\\ \hline
Portugal	&	10.37	&	2.91 \%	&	2.95 \%	&	1.25 \%	\\ \hline
Hungary	&	9.86	&	2.85 \%	&	2.90 \%	&	1.63 \%	\\ \hline
Sweden	&	9.79	&	2.84 \%	&	2.89 \%	&	1.70 \%	\\ \hline
Austria	&	8.58	&	2.70 \%	&	2.77 \%	&	2.64 \%	\\ \hline
Bulgaria	&	7.20	&	2.54 \%	&	2.63 \%	&	3.89 \%	\\ \hline
Denmark	&	5.65	&	2.35 \%	&	2.48 \%	&	5.51 \%	\\ \hline
Finland	&	5.47	&	2.33 \%	&	2.46 \%	&	5.69 \%	\\ \hline
Slovakia	&	5.40	&	2.32 \%	&	2.45 \%	&	5.80 \%	\\ \hline
Ireland	&	4.63	&	2.22 \%	&	2.37 \%	&	6.73 \%	\\ \hline
Croatia	&	4.23	&	2.18 \%	&	2.33 \%	&	7.25 \%	\\ \hline
Lithuania	&	2.92	&	2.02 \%	&	2.20 \%	&	9.13 \%	\\ \hline
Slovenia	&	2.06	&	1.92 \%	&	2.12 \%	&	10.46 \%	\\ \hline
Latvia	&	1.99	&	1.91 \%	&	2.11 \%	&	10.64 \%	\\ \hline
Estonia	&	1.31	&	1.82 \%	&	2.04 \%	&	11.83 \%	\\ \hline
Cyprus	&	0.85	&	1.77 \%	&	1.99 \%	&	12.72 \%	\\ \hline
Luxembourg	&	0.56	&	1.73 \%	&	1.96 \%	&	13.36 \%	\\ \hline
Malta	&	0.43	&	1.71 \%	&	1.95 \%	&	13.68 \%	\\ \hline

\end{tabular}\\
\caption{Indices for Scotland joining the Union after Brexit}\label{tab5}\end{center}
\end{table}

\begin{table}
\begin{center}
\begin{tabular}{|l|c|c|c|c|}\hline
Country & Population & \multicolumn{2}{c|}{Banzhaf Index (\%)}  & Relative\\
& (millions)& with Sweden & without Sweden&Difference\\
\hline &&&&\\[-4.5mm]\hline
Germany	&	81.09	&	10.19 \%	&	11.23 \%	&	10.15 \%	\\ \hline
France	&	66.35	&	8.45 \%	&	9.27 \%	&	9.79 \%	\\ \hline
United Kingdom	&	64.77	&	8.27 \%	&	9.07 \%	&	9.68 \%	\\ \hline
Italy	&	61.44	&	7.91 \%	&	8.66 \%	&	9.44 \%	\\ \hline
Spain	&	46.44	&	6.20 \%	&	6.73 \%	&	8.46 \%	\\ \hline
Poland	&	38.01	&	5.07 \%	&	5.39 \%	&	6.23 \%	\\ \hline
Romania	&	19.86	&	3.78 \%	&	3.93 \%	&	3.80 \%	\\ \hline
Netherlands	&	17.16	&	3.50 \%	&	3.59 \%	&	2.74 \%	\\ \hline
Belgium	&	11.26	&	2.90 \%	&	2.90 \%	&	0.04 \%	\\ \hline
Greece	&	10.85	&	2.86 \%	&	2.85 \%	&	-0.19 \%	\\ \hline
Czech Republic	&	10.42	&	2.81 \%	&	2.80 \%	&	-0.43 \%	\\ \hline
Portugal	&	10.37	&	2.81 \%	&	2.79 \%	&	-0.47 \%	\\ \hline
Hungary	&	9.86	&	2.76 \%	&	2.73 \%	&	-0.77 \%	\\ \hline
Sweden	&	9.79	&	2.75 \%	&		&		\\ \hline
Austria	&	8.58	&	2.63 \%	&	2.58 \%	&	-1.59 \%	\\ \hline
Bulgaria	&	7.20	&	2.49 \%	&	2.42 \%	&	-2.58 \%	\\ \hline
Denmark	&	5.65	&	2.33 \%	&	2.24 \%	&	-3.80 \%	\\ \hline
Finland	&	5.47	&	2.31 \%	&	2.22 \%	&	-3.96 \%	\\ \hline
Slovakia	&	5.40	&	2.30 \%	&	2.21 \%	&	-4.02 \%	\\ \hline
Ireland	&	4.63	&	2.22 \%	&	2.12 \%	&	-4.73 \%	\\ \hline
Croatia	&	4.23	&	2.18 \%	&	2.07 \%	&	-5.12 \%	\\ \hline
Lithuania	&	2.92	&	2.05 \%	&	1.92 \%	&	-6.46 \%	\\ \hline
Slovenia	&	2.06	&	1.96 \%	&	1.82 \%	&	-7.46 \%	\\ \hline
Latvia	&	1.99	&	1.95 \%	&	1.81 \%	&	-7.56 \%	\\ \hline
Estonia	&	1.31	&	1.88 \%	&	1.73 \%	&	-8.41 \%	\\ \hline
Cyprus	&	0.85	&	1.84 \%	&	1.67 \%	&	-9.04 \%	\\ \hline
Luxembourg	&	0.56	&	1.81 \%	&	1.64 \%	&	-9.44 \%	\\ \hline
Malta	&	0.43	&	1.79 \%	&	1.62 \%	&	-9.64 \%	\\ \hline
\end{tabular}\\
\caption{Indices if Sweden leaves the EU with $28$ members}\label{tab6}\end{center}
\end{table}
\begin{table}
\begin{center}
\begin{tabular}{|l|c|c|c|c|}\hline
Country & Population & \multicolumn{2}{c|}{Banzhaf Index (\%)}  & Relative\\
& (millions)& with Estonia & without Estonia &Difference\\
\hline &&&&\\[-4.5mm]\hline
Germany	&	81.09	&	10.19 \%	&	11.20 \%	&	9.84 \%	\\ \hline
France	&	66.35	&	8.45 \%	&	9.26 \%	&	9.67 \%	\\ \hline
United Kingdom	&	64.77	&	8.27 \%	&	9.06 \%	&	9.55 \%	\\ \hline
Italy	&	61.44	&	7.91 \%	&	8.65 \%	&	9.27 \%	\\ \hline
Spain	&	46.44	&	6.20 \%	&	6.73 \%	&	8.51 \%	\\ \hline
Poland	&	38.01	&	5.07 \%	&	5.37 \%	&	5.87 \%	\\ \hline
Romania	&	19.86	&	3.78 \%	&	3.89 \%	&	2.73 \%	\\ \hline
Netherlands	&	17.16	&	3.50 \%	&	3.55 \%	&	1.57 \%	\\ \hline
Belgium	&	11.26	&	2.90 \%	&	2.86 \%	&	-1.41 \%	\\ \hline
Greece	&	10.85	&	2.86 \%	&	2.81 \%	&	-1.68 \%	\\ \hline
Czech Republic	&	10.42	&	2.81 \%	&	2.76 \%	&	-1.94 \%	\\ \hline
Portugal	&	10.37	&	2.81 \%	&	2.75 \%	&	-1.96 \%	\\ \hline
Hungary	&	9.86	&	2.76 \%	&	2.69 \%	&	-2.32 \%	\\ \hline
Sweden	&	9.79	&	2.75 \%	&	2.68 \%	&	-2.36 \%	\\ \hline
Austria	&	8.58	&	2.63 \%	&	2.54 \%	&	-3.22 \%	\\ \hline
Bulgaria	&	7.20	&	2.49 \%	&	2.38 \%	&	-4.31 \%	\\ \hline
Denmark	&	5.65	&	2.33 \%	&	2.20 \%	&	-5.69 \%	\\ \hline
Finland	&	5.47	&	2.31 \%	&	2.17 \%	&	-5.87 \%	\\ \hline
Slovakia	&	5.40	&	2.30 \%	&	2.17 \%	&	-5.93 \%	\\ \hline
Ireland	&	4.63	&	2.22 \%	&	2.07 \%	&	-6.72 \%	\\ \hline
Croatia	&	4.23	&	2.18 \%	&	2.03 \%	&	-7.16 \%	\\ \hline
Lithuania	&	2.92	&	2.05 \%	&	1.87 \%	&	-8.67 \%	\\ \hline
Slovenia	&	2.06	&	1.96 \%	&	1.77 \%	&	-9.79 \%	\\ \hline
Latvia	&	1.99	&	1.95 \%	&	1.76 \%	&	-9.89 \%	\\ \hline
Estonia	&	1.31	&	1.88 \%	&		&		\\ \hline
Cyprus	&	0.85	&	1.84 \%	&	1.62 \%	&	-11.56 \%	\\ \hline
Luxembourg	&	0.56	&	1.81 \%	&	1.59 \%	&	-12.02 \%	\\ \hline
Malta	&	0.43	&	1.79 \%	&	1.57 \%	&	-12.22 \%	\\ \hline

\end{tabular}\\
\caption{Indices if Estonia leaves the EU with $28$ members}\label{tab7}\end{center}
\end{table}

\newpage
\section{Appendix B}

\bgroup
\def\arraystretch{1.5}
	\centering
	\addtolength{\LTleft} {-0.68in}
	\addtolength{\LTright}{-0.68in}
\begin{longtable}{|c|c|c|p{0.84\textwidth}|}
	\hline
	\textbf{$x \in c\,$} & \textbf{pre-exit} & \textbf{post-exit} & \textbf{conditions} \\
	\hline
	$x \notin c$ & $\cL^{28}(v)$ & $\cL^{27}(v)$ & %
		$\#c < 15$ or $w(c) < q_2 (1 - w_x)$ \\
	\hline
	$x \notin c$ & $\cL^{28}(v)$ & $\cW_0^{27}(v)$ & %
		$\#c > 15$ and $q_2 > w(c) > q_2 (1 - w_x) + w_v $ \\
	\hline
	$x \notin c$ & $\cL^{28}(v)$ & $\cW_1^{27}(v)$ & %
		$\#c = 15$ and $w(c) > q_2 (1 - w_x) + w_v $ \\
	\hline
	$x \notin c$ & $\cL^{28}(v)$ & $\cW_2^{27}(v)$ & $\#c > 15$ and %
		$q_2 (1 - w_x) + w_v > w(c) \ge q_2 (1 - w_x)$ \\
	\hline
	$x \notin c$ & $\cL^{28}(v)$ & $\cW_2^{27}(v)$ & $\#c = 15$ and %
		$q_2 (1 - w_x) + w_v > w(c) \ge q_2 (1 - w_x)$ \\
	\hline
	$x \in c$ & $\cL^{28}(v)$ & $\cL^{27}(v)$ & %
		$\#c < 16$ or $w(c) < q_2$ \\
	\hline
	$x \in c$ & $\cL^{28}(v)$ & $\cW^{27}(v)$ & %
		impossible, as $\#c' < \#c <= 15$ \\
	\hline
	\multirow{2}{*}{$x\notin c$} & \multirow{2}{*}{$\cW^{28}(v)$} &  \multirow{2}{*}{$\cL^{27}(v)$} & %
		\multirow{1}{*}{\begin{tabular}[x]{@{}l@{}}
			impossible, as $\#c' = \#c > 16$ and \\
			$w(c') - w(v') = \frac{w(c) - w_v}{1 - w_x} > w(c) - w_v > q_2$
		\end{tabular}} \\
	& & & \\
	\hline
	\multirow{2}{*}{$x\notin c$} & $\cW_0^{28}(v)$ & %
		\multirow{2}{*}{$\cW_0^{27}(v)$} & \multirow{2}{*}{
			$\#c >= 16 > 15$ and $w(c) > q_2 + w_v$
		} \\
	& $\cW_1^{28}(v)$ & & \\
	\hline
	\multirow{1}{*}{$x\notin c$} & $\cW^{28}(v)$ &  \multirow{1}{*}{$\cW_1^{27}(v)$} & %
		\multirow{1}{*}{impossible, as $\#c' = \#c >= 16 > 15$} \\
	\hline
	\multirow{2}{*}{$x\notin c$} & $\cW_0^{28}(v)$ &  \multirow{2}{*}{$\cW_2^{27}(v)$} & %
		\multirow{2}{*}{
			impossible, as $w(c') - w(v') = \frac{w(c) - w_v}{1 - w_x} > w(c) - w_v > q_2$
		} \\
	& $\cW_1^{28}(v)$ & & \\
	\hline
	\multirow{2}{*}{$x\notin c$} & $\cW_0^{28}(v)$ &  \multirow{2}{*}{$\cW_3^{27}(v)$} & %
		\multirow{2}{*}{impossible because of the conjunction of the above two reasons} \\
	& $\cW_1^{28}(v)$ & & \\
	\hline
	\multirow{2}{*}{$x\notin c$} & $\cW_2^{28}(v)$ & \multirow{2}{*}{$\cW_0^{27}(v)$} & %
		\multirow{2}{*}{$q_2 + w_v > w(c) > \max\{q_2 (1 - w_x) + w_v, q_2\} $} \\
	& $\cW_3^{28}(v)$ & & \\
	\hline
	\multirow{2}{*}{$x\notin c$} & $\cW_2^{28}(v)$ & \multirow{2}{*}{$\cW_1^{27}(v)$} & %
		\multirow{2}{*}{impossible, as $\#c' = \#c >= 16 > 15$} \\
	& $\cW_3^{28}(v)$ & & \\
	\hline
	\multirow{2}{*}{$x\notin c$} & $\cW_2^{28}(v)$ & \multirow{2}{*}{$\cW_2^{27}(v)$} & %
		\multirow{2}{*}{$q_2 (1 - w_x) + w_v > w(c) > q_2 > w(c) - w_v $} \\
	& $\cW_3^{28}(v)$ & & \\
	\hline
	\multirow{2}{*}{$x\notin c$} & $\cW_2^{28}(v)$ & \multirow{2}{*}{$\cW_3^{27}(v)$} & %
		\multirow{2}{*}{impossible, as $\#c' = \#c >= 16 > 15$} \\
	& $\cW_3^{28}(v)$ & & \\
	\hline
	$x\in c$ & $\cW_0^{28}(v)$ & $\cL^{27}(v)$ & %
		$q_2 (1 - w_x) + w_x > w(c) > q_2 + w_v $ \\
	\hline
	$x\in c$ & $\cW_0^{28}(v)$ & $\cW_0^{27}(v)$ & %
		$w(c) > q_2 (1 - w_x) + w_x + w_v $ \\
	\hline
	$x\in c$ & $\cW_0^{28}(v)$ & $\cW_1^{27}(v)$ & %
		impossible, as $\#c' = \#c - 1 > 16$, so $\#c' > 15$ \\
	\hline
	$x\in c$ & $\cW_0^{28}(v)$ & $\cW_2^{27}(v)$ & %
		$q_2 (1 - w_x) + w_x + w_v > w(c) > \max\{{q_2 (1 - w_x) + w_x \atop q_2 + w_v}\} $ \\
	\hline
	$x\in c$ & $\cW_0^{28}(v)$ & $\cW_3^{27}(v)$ & %
		impossible, as $\#c' = \#c - 1 > 16$, so $\#c' > 15$ \\
	\hline
	$x\in c$ & $\cW_1^{28}(v)$ & $\cL^{27}(v)$ & %
		$q_2 (1 - w_x) + w_x > w(c) > q_2 + w_v $ \\
	\hline
	$x\in c$ & $\cW_1^{28}(v)$ & $\cW_0^{27}(v)$ & %
		impossible, as $\#c' = \#c - 1 = 15$ \\
	\hline
	$x\in c$ & $\cW_1^{28}(v)$ & $\cW_1^{27}(v)$ & %
		$w(c) > q_2 (1 - w_x) + w_x + w_v $ \\
	\hline
	$x\in c$ & $\cW_1^{28}(v)$ & $\cW_2^{27}(v)$ & %
		impossible, as $\#c' = \#c - 1 = 15$ \\
	\hline
	$x\in c$ & $\cW_1^{28}(v)$ & $\cW_3^{27}(v)$ & %
		$q_2 (1 - w_x) + w_x + w_v > w(c) > \max\{{q_2 (1 - w_x) + w_x \atop q_2 + w_v}\} $ \\
	\hline
	\multirow{2}{*}{$x\in c$} & $\cW_2^{28}(v)$ & \multirow{2}{*}{$\cL^{27}(v)$} & %
		\multirow{2}{*}{$w(c) < \min\{q_2 (1 - w_x) + w_x, q_2 + w_v\} $} \\
	& $\cW_3^{28}(v)$ & & \\
	\hline
	\multirow{2}{*}{$x\in c$} & $\cW_2^{28}(v)$ & \multirow{2}{*}{$\cW_0^{27}(v)$} & %
		\multirow{2}{*}{\begin{tabular}[x]{@{}l@{}}
			impossible, as $w(c) - w_v > q_2 (1 - w_x) + w_x > q_2$ \\
			and $w(c) - w_v < q_2$ are contradictory
		\end{tabular}} \\
	& $\cW_3^{28}(v)$ & & \\
	\hline
	$x\in c$ & $\cW_2^{28}(v)$ & $\cW_1^{27}(v)$ & %
		impossible, as $\#c' = \#c - 1 > 16$, so $\#c' > 15$ \\
	\hline
	$x\in c$ & $\cW_2^{28}(v)$ & $\cW_2^{27}(v)$ & %
		$q_2 + w_v > w(c) > q_2 (1 - w_x) + w_x $ \\
	\hline
	$x\in c$ & $\cW_2^{28}(v)$ & $\cW_3^{27}(v)$ & %
		impossible, as $\#c' = \#c - 1 > 16$, so $\#c' > 15$ \\
	\hline
	$x\in c$ & $\cW_3^{28}(v)$ & $\cW_1^{27}(v)$ & %
		impossible, as $w(c) - w_v > q_2 (1 - w_x) + w_x > q_2$ \newline
		and $w(c) - w_v < q_2$ are contradictory \\
	\hline
	$x\in c$ & $\cW_3^{28}(v)$ & $\cW_2^{27}(v)$ & %
		impossible, as $\#c' = \#c - 1 = 15$ \\
	\hline
	$x\in c$ & $\cW_3^{28}(v)$ & $\cW_3^{27}(v)$ & %
		$q_2 + w_v > w(c) > q_2 (1 - w_x) + w_x $ \\
	\hline
	\caption{Coalitions status changes for country $v$ after a country $x$ exits the EU of 28 and the conditions under which they happen.}
	\label{tabXProd}
\end{longtable}
\egroup

\noindent
\end{document}